\documentclass[lean]{AFM}
\usepackage{amsmath, amsfonts, amsthm, framed}
\usepackage{xspace}
\usepackage{newunicodechar}
\usepackage{enumitem}
\usepackage{fontawesome}
\usepackage{soul}
\usepackage{cancel}
\usepackage{xspace}
\usepackage{eucal}
\usepackage{colonequals}
\definecolor{keywordcolor}{rgb}{0.7, 0.1, 0.1}   % red
\definecolor{commentcolor}{rgb}{0.4, 0.4, 0.4}   % grey
\definecolor{symbolcolor}{rgb}{0, 0, 0.8}    % blue
\definecolor{tacticcolor}{rgb}{0, 0, 0.8}    % blue
\definecolor{sortcolor}{rgb}{0.1, 0.5, 0.1}      % green
\theoremstyle{plain}
\theoremstyle{definition}
\usepackage{stackengine}
\usepackage{wasysym}
\usepackage{enumitem}
\usepackage{booktabs}
\usepackage[english]{babel}
\usepackage{needspace}
\usepackage{isabelle}
\usepackage{isabellesym}
\usepackage{verbatim}
\usepackage{mdframed}
\newcommand{\etal}{\textit{et al.}}
\newcommand{\ie}{\textit{i.e.}}
\newcommand{\Ie}{\textit{I.e.}}
\newcommand{\eg}{\textit{e.g.}}
\newcommand{\suchthat}[0]{\text{s.t.}}
\DeclareMathOperator*{\expectation}{\mathbb{E}}
\DeclareMathOperator{\var}{Var}
\DeclareMathOperator{\prob}{\mathcal{P}}
\DeclareMathOperator{\omod}{\mathrm{mod}}
\newcommand{\bigo}{\mathcal O}
\newcommand{\transpose}[1]{{#1}^\intercal\mkern-2mu}

\title{Derandomization with pseudorandomness}
\author[Emin Karayel]{Emin Karayel}
\authorinfo[Emin Karayel]{Technical University of Munich, School of Computation, Information and Technology, Germany}{me@eminkarayel.de}

\begin{abstract}
Derandomization techniques are often used within advanced randomized algorithms.
In particular, pseudorandom objects, such as hash families and expander graphs, are key components of such algorithms, but their verification presents a challenge.
This work shows how such algorithms can be expressed and verified in Isabelle and presents a pseudorandom objects library that abstracts away the deep algebraic/analytic results involved.
Moreover, it presents examples that show how the library eases and enables the verification of advanced randomized algorithms.
Highlighting the value of this framework is that it was recently used to verify the space-optimal distinct elements algorithm by B\l{}asiok from 2018,
which relies on the combination of many derandomization techniques to achieve its optimality.
\end{abstract}



\VOLUME{1}
\YEAR{2025}
\firstpage{57}
\DOI{https://doi.org/10.46298/afm.13477}
\receiveddate{April 26, 2024}
\finaldate{May 8, 2025}
\accepteddate{May 13, 2025}

\isabellestyle{literal}

\definecolor{shadecolor}{gray}{0.95}
\newenvironment{isabelle_cm}{\begin{mdframed}[innerleftmargin=0.6cm,backgroundcolor=shadecolor,nobreak=true,linewidth=0]\begin{isabelle}}{\end{isabelle}\end{mdframed}}%

\begin{document}

\section{Introduction}
Computer scientists are familiar with pseudorandom number generators (PRNGs), functions that consume a small sequence of seed bits and output a much larger sequence of \emph{pseudorandom} bits.
Well-known examples are the Lehmer random number generator~\cite{oneill2014,payne1969}, Blum Blum Shub~\cite{blum1983}, and the Mersenne-Twister~\cite{matsumoto1998}.
The randomness of these functions can only be verified empirically~\cite{bhattacharjee2018,singh2020}.
But there are no mathematical, in particular, probability-theoretic theorems about them that would make them useful in the design of provably correct randomized algorithms.\footnote{
There are conditional results, \eg, with the assumption of the difficulty of factoring.
Moreover, there are rigorous results about the period of some PRNGs, but they are usually not helpful with respect to the correctness of algorithms using such PRNGs.}

On the other hand, there are \emph{pseudorandom objects}\footnote{The term was coined by Vadhan~\cite{vadhan2012} but it is applied here more generally than he does.}, such as $k$-independent hash families and expander graphs, satisfying such theorems, for example, limited independence or Chernoff-type tail bounds.
They are often critical components of advanced randomized algorithms.
Similar to PRNGs, they are functions taking a small sequence of seed bits and returning a much larger sequence of output bits. Often, the result is represented as a function or a vector.

Pseudorandom objects are used as a derandomization technique and thus improve the randomness usage of simpler algorithms.
This is especially interesting in cases where the saving in coin-flips translates into a reduction of space usage because the algorithm needs to preserve the coin-flips during its operation.
But, as we will see in this work, the construction of pseudorandom objects relies on advanced mathematical foundations, such as finite fields and spectral graph theory.
This makes their formalization in proof assistants a challenge, as well as the verification of randomized algorithms relying on them.
This work shows how advanced randomized algorithms can be expressed and verified in Isabelle/HOL~\cite{nipkow2002} (and in general in every proof assistant), through the use of a composable pseudorandom objects library.

\paragraph{A motivating example:} To illustrate the utility of pseudorandom objects, let us look into the following simple randomized streaming algorithm. For readers less familiar with Isabelle syntax: It is possible to skip these. There is always a precise description of these algorithms in the surrounding text.
\begin{isabelle_cm}
\isacommand{fun}\ example{\isacharunderscore}{\kern0pt}1\ \isacharcolon\isacharcolon\ nat\ \isasymRightarrow\ nat\ list\ \isasymRightarrow\ real\ pmf\isanewline
\ \ \isakeyword{where}\ example{\isacharunderscore}{\kern0pt}1\ n\ xs\ \isacharequal\ \isanewline
\ \ \ \ do\ \isacharbraceleft{\kern0pt}\isanewline
\ \ \ \ \ \ h\ \isasymleftarrow\ prod{\isacharunderscore}{\kern0pt}pmf\ \isacharbraceleft\isadigit{0}\isachardot\isachardot{\isacharless}n\isacharbraceright\ \isacharparenleft\isasymlambda\isacharunderscore\isachardot\ pmf\isacharunderscore{\kern0pt}of\isacharunderscore{\kern0pt}set\ \isacharbraceleft\isacharminus\isadigit{1}\isacharcomma\isadigit{1}\isacharbraceright\isacharparenright\isacharsemicolon\isanewline
\ \ \ \ \ \ return\isacharunderscore{\kern0pt}pmf\ \isacharparenleft\isacharparenleft{\isasymSum}x\ \isasymleftarrow\ xs\isachardot\ h\ x\isacharparenright\isactrlsup {\isadigit{2}}\isacharparenright\isanewline
\ \ \ \ \isacharbraceright
\end{isabelle_cm}
This is a randomized approximation algorithm for the second-frequency moment\footnote{The second frequency moment is the square sum of the occurrence counts of each value in the stream.
It is useful in assessing the skew of the stream, which can, for example, be needed in monitoring applications or query optimization~\cite{dewitt1992,gufler2011}.
The algorithm~\cite{alon1999} is very concise, which makes it useful as an illustrative example.} $F_2$ of a sequence $x_1,\ldots,x_m$ where the sequence values are assumed to be in the set $U := \{0,\ldots,n-1\}$.

In the first step, the algorithm randomly chooses a function $h$ from $U$ (the set containing the sequence values) into the two-element set $\{-1,1\}$, \ie, it assigns each element of $U$ a random signed unit.
The algorithm then returns the square of the sum of the signed units assigned to each sequence element, \ie: $(\sum_{i=1}^{m} h(x_i))^2$.
It is effortless to verify that the algorithm returns an approximation of $F_2$. Indeed, the expected result is:
\begin{equation}
\label{eq:ex1_exp}
\expectation \left(\sum_{i=1}^{m} h(x_i)\right)^2 = \expectation \left(\sum_{u\in U} c(u) h(u)\right)^2 = \sum_{u \in U} \sum_{v \in U} c(u) c(v) \expectation h(u) h(v) = \sum_{u \in U} c(u)^2  \textrm{.}
\end{equation}
Here, $c(u)$ denotes the occurrence count of $u$ in the stream $x_1,\ldots,x_m$.
The key step in the above derivation is the third equality, where the expectation $\expectation h(u) h(v)$ is $0$ if $u \neq v$ and $1$ otherwise.
The second frequency moment $F_2$ is $\sum_{u \in U} c(u)^2$ which corresponds to the right hand side of the equation chain.
A similar argument can be used to derive:
\begin{equation}
\label{eq:ex1_var}
\var \left(\sum_{i=1}^{m} h(x_i)\right)^2 \leq 2 F_2^2 \textrm{.}
\end{equation}
The algorithm above is, however, not very efficient: It requires $\bigo(n + \ln m)$ bits of memory.
At least $n$ bits are required to store the function $h$ and another $\bigo(\ln m)$ bits for the accumulation register while evaluating the sum.
As promised, it can be improved using the ``magic of pseudorandomness''.
Instead of choosing the function $h$ completely randomly, which necessarily requires $\bigo(n)$ bits of storage space, it can be selected using a pseudorandom object with a small seed space.
To achieve that, the function has to be stored implicitly using the seed instead of as a dictionary.

Let us recall that a set of random variables are $k$-wise independent if any $k$ subset of them are independent.
And note that the expectation of a product of independent random variables is the product of their expectations.
Indeed, a careful examination of Eq.~\ref{eq:ex1_exp}, shows that the values of the function $h$ only have to be pair-wise independent.
In Eq.~\ref{eq:ex1_exp} in the last step, we evaluate the product of only pairs of the random variables $h(u)$, $h(v)$.
A similar observation can be made for Eq.~\ref{eq:ex1_var}, where $4$-wise independence is sufficient.

This means that the pseudorandom object we want is required to form a $4$-wise independent family of functions.
The following example is the formalization of the updated algorithm:
\begin{isabelle_cm}
\isacommand{fun}\ example{\isacharunderscore}{\kern0pt}2\ \isacharcolon\isacharcolon\ nat\ \isasymRightarrow\ nat\ list\ \isasymRightarrow\ real\ pmf\isanewline
\ \ \isakeyword{where}\ example{\isacharunderscore}{\kern0pt}2\ n\ xs\ \isacharequal\ \isanewline
\ \ \ \ do\ \isacharbraceleft{\kern0pt}\isanewline
\ \ \ \ \ \ h\ \isasymleftarrow\ sample{\isacharunderscore}{\kern0pt}pro\ \isacharparenleft\isasymH\ \isadigit{4}\ n\ \isacharparenleft\isasymL\ \isacharbrackleft\isadigit{1}\isacharcomma\isacharminus\isadigit{1}\isacharbrackright\isacharparenright\isacharparenright\isanewline
\ \ \ \ \ \ return\isacharunderscore{\kern0pt}pmf\ \isacharparenleft\isacharparenleft{\isasymSum}x\ \isasymleftarrow\ xs\isachardot\ h\ x\isacharparenright\isactrlsup {\isadigit{2}}\isacharparenright\isanewline
\ \ \ \ \isacharbraceright
\end{isabelle_cm}
The new algorithm has the same properties as \isa{example{\isacharunderscore}1} (\ie, satisfying Eq.~\ref{eq:ex1_exp} and \ref{eq:ex1_var}) but requires only $\bigo(\ln n + \ln m)$ bits of memory.
The only difference is in Line~4, where $h$ is now obtained using the pseudorandom object specified by the term \isa{\isasymH\ \isadigit{4}\ n\ \isacharparenleft\isasymL\ \isacharbrackleft\isadigit{1}\isacharcomma\isacharminus\isadigit{1}\isacharbrackright\isacharparenright}, which requires $\bigo(\ln n)$ bits of input as seed and returns a function from $\{0,\ldots,n-1\}$ to $\{-1,1\}$ (which is stored implicitly using the $\bigo(\ln n)$ bits of space, instead of as a dictionary).
If the seed is chosen uniformly at random, then the evaluation of the function at any point in $\{0,\ldots,n-1\}$ forms a random variable.
Crucially, the family of all $n$ random variables is $4$-wise independent, and each is distributed uniformly on $\{-1,1\}$.
The operation \isa{sample{\isacharunderscore}pro} interprets a pseudorandom object as a probability space.
It uniformly chooses a random seed from the seed space of the pseudorandom object and evaluates the object at that seed. 

The expression \isa{\isasymH\ \isadigit{4}\ n\ \isacharparenleft\isasymL\ \isacharbrackleft\isadigit{1}\isacharcomma\isacharminus\isadigit{1}\isacharbrackright\isacharparenright} from above represents a combined pseudorandom object. 
It is defined using two constructs: The expression \isa{\isasymL\ \isacharbrackleft\isadigit{1}\isacharcomma\isacharminus\isadigit{1}\isacharbrackright} describes a pseudorandom object with a single-bit input that outputs $-1$ or $1$ depending on the bit.
The construct \isa{\isasymH\ k\ n\ P} is a $k$-independent hash family, with domain $\{0,\ldots,n-1\}$, and the range $\isa{P}$, where \isa{P} is another pseudorandom object.
This means it is easy to construct application-specific pseudorandom objects from a small set of primitives.
The presented ones are part of the libraries~\cite{Universal_Hash_Families-AFP, Expander_Graphs-AFP} (available in the AFP~\cite{afp} for Isabelle.)

The step from \isa{example{\isacharunderscore}1} to \isa{example{\isacharunderscore}2} demonstrates derandomization with pseudorandom objects.
The above is just a teaser example and will be further expanded on in Section~\ref{sec:combining_pros}, where combinations of multiple types of pseudorandom objects, in particular, including expander graphs, will take the stage.

\paragraph{Contributions:} The main novel results presented in this work are:
\begin{itemize}
\item A novel approach to constructing pseudorandom objects based on simple primitives and the introduction of notation for them that allows a succinct representation within algorithms using them.
\item Formal verification of an efficient and effective library for computations in both prime and non-prime finite fields, \ie, $GF(p^n)$ for $p$ prime and any $n \geq 1$.
Prior work in proof assistants~\cite{divason2020, Berlekamp_Zassenhaus-AFP} focused on the much simpler case of prime finite fields, \ie, the $n=1$ case.
This is the basis for the first pseudorandom object---$k$-independent hash families---detailed out in Section~\ref{sec:hash}.
\item Formal verification of expander graphs and a construction of a strongly explicit family for every degree and spectral bound and the verification of tail bounds for random walks in expander graphs.
This is the basis for the second pseudorandom object---random walks in expander graphs---detailed out in Section~\ref{sec:expander}.
\end{itemize}
It should be noted that the examples are verified~\cite[Tutorial{\textunderscore}Pseudorandom{\textunderscore}Objects]{Frequency_Moments-AFP} in Isabelle.
The supporting libraries require more than 16300 lines of code\footnote{Appendix~\ref{apx:form} provides a comprehensive list of the formalizations mentioned here.} because the construction of the pseudorandom objects builds on mathematical foundations such as (non-prime) finite fields and spectral graph theory.
However, the libraries can be used easily based on high-level results that abstract away implementation details.

The following sections will discuss related work and additional important background.
The formalization of pseudorandom objects in an abstract way is detailed out in Section~\ref{sec:pros}, followed by the concrete pseudorandom objects $k$-independent hash families (in Section~\ref{sec:hash}) and expander graphs (in Section~\ref{sec:expander}).
Section~\ref{sec:combining_pros} presents an application of the framework, where multiple types of pseudorandom objects are going to combine to improve the previous introductory example.
Section~\ref{sec:concl} concludes with a summary and future research opportunities.

\section{Related work}
\subsection{Derandomization with pseudorandomness}
As mentioned in the abstract, an exciting use case for this framework is the distinct elements problem---estimating the number of distinct elements in a stream in a single pass with limited mutable state.
The research on the problem started in 1985 by Flajolet~\cite{flajolet1985}, followed by successive improvements~\cite{alon1999, baryossef2002, kane2010} until an optimal solution was found by B\l{}asiok in 2018~\cite{blasiok2018,blasiok2020}.
In 2023~\cite{karayel2023,Distributed_Distinct_Elements-AFP,karayel2023arxiv}, I wrote about its formalization using the framework for pseudorandom objects presented here.
It is an improved version of B\l{}asiok's solution. It has smaller implementation complexity, because it uses fewer kinds of pseudorandom objects, and supports a merge operation making the algorithm parallelizable, while still maintaining the optimal space-complexity. %
The concrete pseudorandom object used there is an expander random walk chained with a second expander random walk chained with a product of pseudorandom objects for $k$-independent hash families.
Indeed, the formalization of this algorithm was a big part of the motivation for the design of the library presented in this work.
At the current state of research, pseudorandom objects are essential for the distinct elements problem.
A notable exception is the solution by Chakraborty \etal~\cite{chakraborty2022} that does not require pseudorandom objects, but its space complexity is asymptotically worse than the optimal solution.

There are, of course, many further randomized algorithms that rely on pseudorandom objects, such as heavy hitters~\cite{charikar2004}, frequency moments~\cite{alon1999} and other stream statistics~\cite{blasiok2017}, graph matching~\cite{chen2021}, graph sparsification~\cite{czumaj2021} and low-distortion-embeddings~\cite{houen2023}.

Another motivating influence for this work is Goldreich’s chapter~\cite{goldreich2011} on samplers: Algorithms that probe a function at select points and obtain an estimate for the average of the function.
The performance of samplers is measured with respect to the number of probing points, the number of random bits consumed, and the quality of the estimate.
Goldreich also relies mainly on the two fundamental pseudorandom objects mentioned here: expander graphs and hash families.
In contrast to this work, he does not treat pseudorandom objects as first-class objects but the samplers, which use pseudorandom objects as a component.
However, that viewpoint prevents some of the applications I describe here, including the example from the introduction.
Conversely, samplers are easily represented with the framework presented here.

An ancestor of the notation developed here for pseudorandom objects was the work of Kane \etal~\cite{kane2010}, who introduced a notation for hash families and forming products.
However, they do not consider chaining pseudorandom objects, for which there is no prior art.

Vadhan~\cite{vadhan2012} wrote a monograph and survey on pseudorandomness.
As far as I can tell, he coined the term \emph{pseudorandom object}, and his work stands out by covering a large number of derandomization techniques and highlighting connections and differences.
Important to note is that in this work, the term pseudorandom object is used more broadly than in Vadhan’s monograph.

\subsection{Verification of foundational libraries in proof Assistants}
Vilhena and Paulson~\cite{vilhena2020} formalized foundational results on field theory, in particular, that every field admits an algebraically closed extension.
Although their work does not touch on finite fields, and, in particular, there is no emphasis on efficient implementation of the field operations.
The efficient and effective finite fields library in this work~\cite{Finite_Fields-AFP} builds on the theoretical results and framework developed by Vilhena and Paulson.
The mathlib library~\cite{mathlib} in Lean has a proof of existence for both prime and non-prime finite fields.
However, in contrast to this work, they show the existence in a non-constructive way, without providing and verifying algorithms that allow efficient computations in these fields.
A similar non-efficient existence proof has been verified in Mizar~\cite{schwarzweller2024}.

There is work on the formalization of Markov Chains, for example the Perron--Frobenius theorem~\cite{thiemann2021,Stochastic_Matrices-AFP}, and Markov decision processes~\cite{hoelzl2017,schaeffeler2023}. I could not find any work on the formalization of expander graphs.

\subsection{Randomized Algorithms in Proof Assistants}
The best way for the verification of randomized algorithms in Isabelle/HOL is based on the Giry monad~\cite{giry1982}, which is the method applied in this work.
Eberl \etal~\cite{eberl2020, eberl2015} have done foundational research and verified several randomized algorithms demonstrating the feasibility of the approach.
It should however be noted that this is still an active research area~\cite{audebaud2009, bao2021, chatterjee2018, hoelzl2016, lochbihler2016, sun2023}.
Similarly there is work towards transferring the approach to other proof assistants~\cite{tassarotti2018, tassarotti2021}.

There is, however, not yet much research on derandomization techniques (even $k$-independent hash families).
The only research I know of is the formalization of an approximate model counting algorithm that relies on the construction of a $3$-universal XOR-based hash-family by Tan \etal~\cite{Approximate_Model_Counting-AFP,yongkiam2024} in Isabelle/HOL.
Their work uses some libraries I developed during this work.
Conversely, Yong Kiam Tan made several, much appreciated, improvements to my AFP entry on the median method~\cite{Median_Method-AFP}.

In 2021~\cite{karayel2022}, I presented the formalization of randomized algorithms for frequency moments.
It only uses hash families (and only based on prime finite fields) and does not use other pseudorandom objects or combinations of these.
Moreover, the focus of this work is on derandomization techniques, as opposed to the verification of concrete algorithms.

\subsection{Other derandomization methods}
Besides the use of pseudorandom objects, there are also other derandomization methods. The two most prominent are:
\begin{itemize}
\item Non-uniform (non-constructive) derandomization
\item Method of conditional expectations (or probabilities)
\end{itemize}
The first---\emph{non-uniform derandomization}---is based on Adleman's theorem~\cite{adleman1978}, who showed that any randomized algorithm with a finite set $S$ of possible inputs can be derandomized entirely.
This works by first amplifying it until the failure probability is smaller than $\frac{1}{|S|}$.
For such an algorithm it is possible to show that there must exist a fixed choice of coin flips, for which the algorithm must succeed for every input. This follows from a union-bound.
The result is sometimes abbreviated as: $\mathrm{RP} \subseteq \mathrm{P/poly}$.
The theorem does not provide a way to find such a sequence of coin-flips.
The topic is less relevant for formal verification: Constructing such an algorithm would require a brute-force search, and it is trivial to verify the correctness of an algorithm with a finite input set.

The second---\emph{method of conditional expectations}~\cite{alon2000,vadhan2012}---can be applied in some specific cases.
Given a randomized algorithm, it is sometimes possible to derive a deterministic algorithm that performs at least as well as the randomized algorithm on average.
This requires that the probability space can be decomposed recursively into a search tree, where conditional expectations can be computed for each node.
I have formalized two algorithms based on this technique~\cite{Derandomization_Conditional_Expectations-AFP}, but it turned out that it is straightforward to verify such algorithms.
This contrasts pseudorandom objects, which require large foundational libraries to work with them.

\section{Background on randomized algorithms}
This section provides a very small introduction, into the representation of (randomized) algorithms in Isabelle.

Isabelle is a meta-logic for which several instantiations exist~\cite[Part III]{paulson1994}.
The most prominent, and the framework used in this work, is HOL. Other instantiations are, for example, First-order logic (FOL) or Zermelo--Fraenkel Set Theory (ZF).
HOL can be described as a typed version of classical mathematical logic.\footnote{In my opinion HOL is actually closer to informal mathematics used today than, \eg, Isabelle/ZF. For example a textbook proof would never consider the equality of a group and a neighborhood basis of a topology; they are, informally, considered different types of objects, even though formally everything is a set.}

The type system is similar to that of many functional programming languages, such as Haskell or ML.
The primitives \isa{nat}, \isa{real} represent natural (resp.\ real) numbers.
It is also possible to construct more complex types from simple ones, for example \isa{nat\ \isasymRightarrow\ real} represents a function from natural to real numbers.
A bit unexpected is that type constructors are written post-fix, \eg, \isa{nat list} represents a list of natural numbers.
Type variables are denoted using primed letters, \eg: \isa{{\isacharprime}a}, \isa{{\isacharprime}b}.
The double colon is used to signify membership of an expression or name in a particular type, \eg: \isa{n\ \isacharcolon\isacharcolon\ nat}.

Isabelle's expressions resemble standard mathematical notation closely, but there are sometimes slight deviations. The following table summarizes such cases:

\medskip

\noindent\begin{tabular}{l l l}
\toprule
Description & Isabelle Term & Classical Math \\
\midrule
Cardinality of a set $S$ & \isa{card\ S} & $|S|$ \\
Summation over a list \isa{xs\ \isacharequal\ {\isacharbrackleft}x\isactrlsub{1}\isacharcomma\isachardot{\isachardot}{\isacharcomma}x\isactrlsub{m}\isacharbrackright} & \isa{\isacharparenleft{\isasymSum}x \isasymleftarrow\ xs\isachardot\ f x\isacharparenright} & $\sum_{i=1}^{m} f(x_i)$ \\
Summation over indices & \isa{\isacharparenleft{\isasymSum}i\ \isacharless\ n\isachardot\ f i\isacharparenright} & $\sum_{i=0}^{n-1} f(i)$\\
Interval with strict bounds & \isa{{\isacharbraceleft}a\isacharless\isachardot\isachardot{\isacharless}b\isacharbraceright} & $\{x \mid a < x < b\}$\\
Interval with inclusive bounds & \isa{{\isacharbraceleft}a\isachardot{\isachardot}b\isacharbraceright} & $\{x \mid a \leq x \leq b\}$ \\
Strictly upper bounded set & \isa{\isacharbraceleft\isachardot{\isachardot}{\isacharless}b\isacharbraceright} & $\{x \mid x < b\}$\\
\bottomrule
\end{tabular}

\medskip 

{\noindent}The interval notation is flexible. The above are just examples, but any combination of unbounded, strict or inclusive lower and upper bounds are possible.

An important type constructor for this work is the one describing probability mass functions (PMFs): \isa{pmf}.
They are used to represent finite/discrete probability spaces and form a special class of general probability spaces, with the condition that the $\sigma$-algebra is the powerset of the universe of the type. Eberl~\cite{eberl2020} provides a more thorough introduction.
This greatly reduces the difficulty of working with them because measurability becomes trivial.
Because distributions of randomized algorithms will always be supported on a countable set, the condition on the $\sigma$-algebra is not a restriction.
For example \isa{real pmf} represents a discrete probability space whose elements are real numbers.
The following table provides a brief overview of the terms that can be used to construct PMFs.

\medskip

\noindent\begin{tabular}{l p{11cm}}
\toprule
Term & Description \\
\midrule
\isa{pmf{\isacharunderscore}of{\isacharunderscore}set\ S} & For a finite set $S$, the uniform probability space on $S$. (Every element of $S$ is equiprobable.) \\
\isa{pmf{\isacharunderscore}of{\isacharunderscore}multiset\ S} & For a multiset $S$, the probability space on the elements of $S$, with the probability of elements being proportional to their multiplicity. \\
\isa{pair{\isacharunderscore}pmf\ A\ B} & Given probability spaces $A$ and $B$, the probability space supported on the product of the supports of $A$ and $B$. (With each pair $(x,y)$ having the probability of the product of the probabilities of $x$ and $y$.) \\
\isa{prod{\isacharunderscore}pmf\ I\ M} & Finite product of multiple probability spaces. The elements are maps with a domain $I$, the index set, and values in $M(i)$ (depending on the index). \\
\isa{map{\isacharunderscore}pmf\ f\ A} & The probability space representing the distribution of the random variable $f$ over the probability space $A$. \\
\isa{return{\isacharunderscore}pmf\ x} & The probability space of the singleton $\{x\}$. \\
\isa{bind{\isacharunderscore}pmf p f} & This represents algorithmically the composition of two randomized algorithms $p$ and $f$, where $f$ is executed using the result of $p$. Mathematically, \isa{bind{\isacharunderscore}pmf p f} is the probability space for which an event $E$ has the probability $\prob_{\isa{bind{\isacharunderscore}pmf p f}} E := \int \prob_{f x} (E) \, d p(x)$. 

Note also that: \isa{map{\isacharunderscore}pmf\ f\ p\ \isacharequal\ bind{\isacharunderscore}pmf\ p\ {\isacharparenleft}return{\isacharunderscore}pmf\ \isasymcirc\ f\isacharparenright} \\
\bottomrule
\end{tabular}

\medskip

{\noindent}The functions \isa{bind{\isacharunderscore}pmf} and \isa{return{\isacharunderscore}pmf} form the Giry monad for probability mass functions,
because of that, randomized algorithms can be represented more intuitively using the do-notation.
As an example  \isa{bind{\isacharunderscore}pmf\ p\ {\isacharparenleft}{\isasymlambda}x\isachardot\ return{\isacharunderscore}pmf\ {\isacharparenleft}q\ x\isacharparenright\isacharparenright} can be written as:
\begin{isabelle_cm}
do\ \isacharbraceleft\isanewline
\ \ x\ \isasymleftarrow\ p\isacharsemicolon\isanewline
\ \ return{\isacharunderscore}pmf (q x)\isanewline
\isacharbraceright
\end{isabelle_cm}
which represents the distribution of a randomized algorithm that obtains \isa{x} from the probability space \isa{p} and returns \isa{q x}.

\section{Pseudorandom objects~\label{sec:pros}}
This section is less exciting, but it is nevertheless necessary to tackle the basic definitions before discussing the advanced objects.
Let us take a brief look at the representation of pseudorandom objects in Isabelle~\cite[Pseudorandom{\textunderscore}Objects]{Universal_Hash_Families-AFP}.
This is done using a record:
\begin{isabelle_cm}
\isacommand{record}\ {\isacharprime}a\ pseudorandom{\isacharunderscore}object\ \isacharequal\isanewline
\ \ pro{\isacharunderscore}last\ \isacharcolon\isacharcolon\ nat\isanewline
\ \ pro{\isacharunderscore}select\ \isacharcolon\isacharcolon\ nat\ \isasymRightarrow\ {\isacharprime}a
\end{isabelle_cm}
Note that the record type \isa{pseudorandom{\isacharunderscore}object} is parameterized over the type of the elements of the pseudorandom object \isa{{\isacharprime}a}.
A record in Isabelle can be thought of as a tuple type, but with the advantage that the components can be accessed using named selectors. In this case, these are: \isa{pro{\isacharunderscore}last}, \isa{pro{\isacharunderscore}select}.

The selector \isa{pro{\isacharunderscore}last} represents the object's size minus $1$.
The selector \isa{pro{\isacharunderscore}select} is a function that returns an element from the object, given an index between $0$ and \isa{pro{\isacharunderscore}last} (inclusive).
Note that this means that pseudorandom objects always have at least one element.
Effectively, a pseudorandom object is just a non-empty multiset from which it is easy to sample.
The fact that empty pseudorandom objects are disallowed comes in handy because this means that we can associate with any pseudorandom object: a probability space.
The function \isa{sample{\isacharunderscore}pro} does exactly that.
It is defined as:
\begin{isabelle_cm}
sample{\isacharunderscore}pro\ P\ =\ map{\isacharunderscore}pmf\ {\isacharparenleft}pro{\isacharunderscore}select\ P\isacharparenright\ {\isacharparenleft}pmf{\isacharunderscore}of{\isacharunderscore}set\ {\isacharbraceleft}\isadigit{0}\isachardot{\isachardot}pro{\isacharunderscore}last\ P{\isacharbraceright}\isacharparenright
\end{isabelle_cm}
While we will keep explicitly using the term \isa{sample{\isacharunderscore}pro}, it is actually defined as a coercion. Whenever a pseudorandom object is used in a place that requires a probability space, Isabelle would automatically convert it to a probability space by introducing an application of \isa{sample{\isacharunderscore}pro}. This is similar to the automatic coercion of integers to real numbers.

Note that the function \isa{pro{\isacharunderscore}select} is not required to be injective.
Such a condition would incur a performance cost for the more complex pseudorandom objects, such as expander graphs.
For an element $x$ of a pseudorandom object, we call the count of indices in \isa{{\isacharbraceleft}0\isachardot{\isachardot}pro{\isacharunderscore}size\ P\isacharbraceright} mapped to $x$ the multiplicity of $x$.

The following are trivial pseudorandom objects and combinators defined in the AFP library~\cite[Pseudorandom{\textunderscore}Objects]{Universal_Hash_Families-AFP}:
\begin{itemize}
\item If \isa{xs} is a non-empty list, then \isa{\isasymL\ xs} is a pseudorandom object. The multiplicity of an element is its occurrence count in the list. In particular: \isa{sample{\isacharunderscore}pro\ \isacharparenleft\isasymL\ xs\isacharparenright\ \isacharequal\ pmf{\isacharunderscore}of{\isacharunderscore}multiset\ {\isacharparenleft}mset\ xs\isacharparenright}.
\item If $n > 0$, then \isa{\isasymN\ n} is the pseudorandom object for the natural numbers $\{0,\ldots,n-1\}$ (each with multiplicity $1$). In particular: \isa{sample{\isacharunderscore}pro\ \isacharparenleft\isasymN\ n\isacharparenright\ \isacharequal\ pmf{\isacharunderscore}of{\isacharunderscore}set\ \isacharbraceleft\isadigit{0}\isachardot\isachardot{\isacharless}n\isacharbraceright}.
\item If \isa{P} and \isa{Q} are pseudorandom objects, then \isa{P\ \isasymtimes\isactrlsub P Q} denotes the product pseudorandom object.
Its size is the product of the sizes of \isa{P} and \isa{Q} and consists of pairs whose first component is from P and whose second component is from $Q$.
The multiplicity of a pair $(x,y)$ is the product of the multiplicities of $x$ in $P$ and $y$ in $Q$. \\ In particular: 
\isa{sample{\isacharunderscore}pro\ {\isacharparenleft}P\ \isasymtimes\isactrlsub P Q\isacharparenright\ \isacharequal\ pair{\isacharunderscore}pmf\ {\isacharparenleft}sample{\isacharunderscore}pro\ P\isacharparenright\ {\isacharparenleft}sample{\isacharunderscore}pro\ P\isacharparenright}.
\end{itemize}
In the following two sections, we will dive into the more exciting pseudorandom objects. The following section is dedicated to $k$-independent hash families and Section~\ref{sec:expander} to random walks in expander graphs.

\section{Hash families~\label{sec:hash}}
The textbook~\cite[3.31]{vadhan2012} definition of $k$-independent hash families is as a family of functions from a finite domain $D$ to a finite range $R$.
For any domain element $d \in D$, we can associate a random variable $X_d$---the evaluation of the function at the point $d$, where the probability space is the hash family.
The key condition $k$-independent hash families satisfy is that the random variables $X_d$ are $k$-wise independent and uniformly distributed in the range $R$.
More formally, the condition can be expressed as:
\[
  \prob \left( \bigwedge_{s \in S} X_s \in A_s \right) = \prod_{s \in S} \prob( X_s \in A_s) = \frac{|A_s|}{|R|}
\]
for all subsets $S \subseteq D$ with at most $k$ elements and arbitrary sets $A_s \subseteq R$.

\begin{figure}[ht!]
\centering
\includegraphics{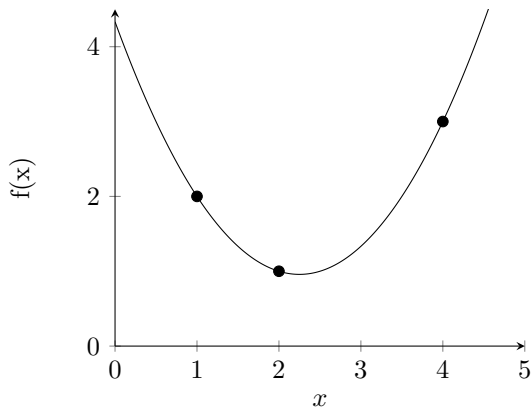}
\caption{There is exactly one polynomial of degree less than 3 interpolating 3 points on the real line. The same is true in finite fields.}
\label{fig:interpol}
\end{figure}

The classic construction for such families is due to Wegman and Carter~\cite{wegman1981}, where the hash functions are just polynomials (with degrees strictly smaller than k) over finite fields. It is straightforward to verify the $k$-independence, which follows from the algebraic fact that there is exactly one polynomial (of degree smaller than $k$) interpolating a given set of $k$ points. (See also Figure~\ref{fig:interpol}.)

In particular, using the Galois field $\mathrm{GF}(q)$, where $q$ is a prime power, we obtain a family 
of functions on $[q] \rightarrow [q]$.
Here, the notation $[n]$ stands for the set of the first $n$ natural numbers, ie, $[n] := \{0,\ldots,n-1\}$.
The seed space for the hash family is $[q^k]$, which is converted to the coefficients of a degree $k$ polynomial.

It is straightforward to construct $k$-independent hash families with different domains and ranges; for example, if $m$ divides $q$, then composition with $x \mapsto x \omod m$ does not affect the $k$-wise independence or uniform distribution property.
This allows us to construct hash families from $[q]$ to $[m]$, if $m$ divides $q$ and $q$ is a prime power using the Galois field $\mathrm{GF}(q)$. Note that $m$ must itself be a prime power, since any divisor of a prime power is again a prime power.
Moreover, we of course do not have to use the entire domain $[q]$, \ie, we can regard the hash family from $[q]$ to $[m]$ as a hash family from $[n]$ to $[m]$, if $[n] \subseteq [q]$ or equivalently $n \leq q$.
This means we can construct a $k$-independent family for functions between any domain of size $n$ and range of size $m$, where $n$ is arbitrary, and $m$ must be a prime power. This works by choosing $q$ as $p^{j}$ where $j := \max(\log_p n, \log_p m)$ and $p$ is the unique prime dividing $m$.

The approach taken in the formalization~\cite[Pseudorandom{\textunderscore}Objects{\textunderscore}Hash{\textunderscore}Families]{Universal_Hash_Families-AFP} goes beyond that because hash families may be used to select the seed of another pseudorandom object.
This is key to the compositionality of the pseudorandom objects library.
Instead of selecting a finite set for the range, in the formalization, the range space can be any other pseudorandom object (if its seed space's size is a prime power).
Internally, this just represents composition with the selection function of the new pseudorandom object.
Notationally, this is represented by the expression \isa{\isasymH\ k\ n\ P} where \isa{P} must be a pseudorandom object with a seed space whose size is a prime power.
The above is not a restriction because we can, of course, just write: \isa{\isasymH\ k\ n\ {\isacharparenleft}\isasymN\ m\isacharparenright} to obtain a family from $[n]$ to $[m]$ (if $m$ is a prime power.)

Note that the seed space of the Carter--Wegman hash family is always a prime power. This means it is possible to chain hash families.  

The defining property of $k$-independent hash families is expressed in the following way:
\begin{isabelle_cm}
map{\isacharunderscore}pmf\ {\isacharparenleft}restrict\ J\isacharparenright\ {\isacharparenleft}sample{\isacharunderscore}pro\ \isacharparenleft\isasymH\ k\ n\ P\isacharparenright\isacharparenright\ \isacharequal\ prod{\isacharunderscore}pmf\ J\ {\isacharparenleft}\isasymlambda{\isacharunderscore}\isachardot\ sample{\isacharunderscore}pro\ P\isacharparenright
\end{isabelle_cm}
under the condition that \isa{J\ \isasymsubseteq\ \isacharbraceleft\isadigit{0}\isachardot\isachardot{\isacharless}n\isacharbraceright} and \isa{card\ J\ \isasymle\ k}. \Ie, if the domains are restricted to any $k$ element set, the functions from the pseudorandom object behave as if they are random functions. More precisely, they behave as if they are sampled independently using the inner pseudorandom object \isa{P}.

The fact that the right-hand side is expressed with respect to the probability space of the chained pseudorandom object \isa{sample{\isacharunderscore}pro\ P} permits the derivation of results in the case of more complex combined pseudorandom objects. For interested readers, Section~\ref{sec:combining_pros} contains a step-by-step proof using combinations of pseudorandom objects.  

\subsection{Construction of finite fields}
Critical to the construction of the Carter--Wegman hash families is the construction of finite fields.
For any prime $p$, the factor ring $\mathrm{GF}(p) := \mathbb{Z}/p\mathbb{Z}$ is a finite field, but not all finite fields can be constructed in this way.
The finite fields of order $p^k$, where $k > 1$, can be constructed using factors of polynomial rings.
In particular, if $f$ is an irreducible polynomial of degree $k$ with coefficients in a finite field $F$, then the factor ring $F[X]/ f F[X]$ becomes a finite field of order $|F|^k$.
This means the construction of the finite field $\mathrm{GF}(p^k)$ reduces to finding an irreducible polynomial of degree $k$ with coefficients in $\mathrm{GF}(p)$.

The latter is helped by the fact that the probability of a random polynomial of degree $k \geq 1$ being irreducible is at least $\frac{1}{2k}$---a consequence of Gauss’s formula counting the number of irreducible polynomials of a given degree over a finite field $\mathrm{GF}(p)$:
\[
\frac{1}{k} \left(\sum_{d | k} \mu(d) p^{k/d}\right)	
\]
In the above formula $\mu(d)$ denotes the M\"{o}bius function, whose value is $0$ if $d$ has a square factor, and otherwise it is $1$ (resp.\ $-1$) if $d$ has an even (resp.\ odd) number of prime factors.
The formula follows from a deep result by Gauss, showing that in a finite field $F$ an irreducible polynomial divides $X^{|F|^n}-X$ if its degree divides $n$, as well as the fact that $X^{|F|^n}-X$ is square-free.

Another consequence of the above is Rabin's irreducibility test~\cite{rabin1980}; a polynomial $f$ is irreducible if it divides $X^{|F|^{\mathrm{deg}(f)}}-X$, but is coprime w.r.t.\ $X^{|F|^{\mathrm{deg}(f)/k}}-X$ for any prime $k$ dividing $\mathrm{deg}(f)$.

The latter suggests a practical algorithm to obtain irreducible polynomials by sampling polynomials randomly and checking for irreducibility. And indeed, this is a practical strategy. The algorithm's expected iteration count is $\bigo(k)$. As far as I know, there are no deterministic algorithms that have a similar performance~\cite{adleman1986, shoup1988}.

The AFP entry~\cite{Finite_Fields-AFP} contains a formalization of these results, in particular, a complete characterization of the finite fields, \ie, that every finite field is of prime power order and finite fields with the same order are isomorphic~\cite{ireland1982,lidl1986,mullen2013}, Gauss formula, the Frobenius homomorphism and Rabin's irreducibility test.
It also introduces efficient algorithms for field operations, \eg, computing the inverse using the extended Euclidean algorithm, as well as long division and fast binary modular exponentiation. 

For interested readers, the following code fragment is the sampling algorithm for irreducible polynomials~\cite[Find{\textunderscore}Irreducible{\textunderscore}Poly]{Finite_Fields-AFP}:
\begin{isabelle_cm}
\isacommand{partial{\isacharunderscore}function}\ {\isacharparenleft}smpf\isacharparenright\ sample{\isacharunderscore}irreducible{\isacharunderscore}poly\ \isacharcolon\isacharcolon\ nat\ \isasymRightarrow\ nat\ \isasymRightarrow\ {\isacharparenleft}nat\ list\ \isasymtimes\ nat\isacharparenright\isanewline
\ \ \isakeyword{where}\ sample{\isacharunderscore}irreducible{\isacharunderscore}poly\ p\ n\ \isacharequal\isanewline
\ \ \ \ do\ \isacharbraceleft\isanewline
\ \ \ \ \ \ k\ \isasymleftarrow\ spmf{\isacharunderscore}of{\isacharunderscore}set\ \isacharbraceleft\isachardot{\isachardot}{\isacharless}p{\isacharcircum}k\isacharbraceright\isacharsemicolon\isanewline
\ \ \ \ \ \ let\ poly\ \isacharequal\ enum{\isacharunderscore}monic{\isacharunderscore}poly\ {\isacharparenleft}mod{\isacharunderscore}ring\ p\isacharparenright\ n\ k\isacharsemicolon\isanewline
\ \ \ \ \ \ if\ rabin{\isacharunderscore}test\ {\isacharparenleft}mod{\isacharunderscore}ring\ p\isacharparenright\ poly\isanewline
\ \ \ \ \ \ \ \ then\ return{\isacharunderscore}spmf\ {\isacharparenleft}poly\isacharcomma\isadigit{1}\isacharparenright\isanewline
\ \ \ \ \ \ \ \ else\ tick{\isacharunderscore}spmf\ {\isacharparenleft}sample{\isacharunderscore}irreducible{\isacharunderscore}poly\ p\ n\isacharparenright\isanewline
\ \ \ \ \isacharbraceright
\end{isabelle_cm}
Note that the algorithm does not terminate unconditionally, but only almost surely, \ie, the probability that it does not terminate is $0$.
This is why the sampling algorithm is represented as a partial function in the SPMF monad. Such functions have a different semantics.
In particular ordinary induction proofs can not be applied but must be replaced with a probabilistic convergence proof.
This relies on the framework by Lochbihler~\cite{Probabilistic_While-AFP}. The entry also contains several illustrative examples.

The algorithm \isa{example{\isacharunderscore}2} from the introduction used a deterministic construction of the finite fields.
(Internally, it chooses the lexicographically smallest irreducible polynomial of a given degree.)
However, as we just saw, a randomized selection of the irreducible polynomial is more efficient.
To use that version, the example needs to be modified slightly:

\needspace{50pt}
\begin{isabelle_cm}
\isacommand{fun}\ example{\isacharunderscore}{\kern0pt}3\ \isacharcolon\isacharcolon\ nat\ \isasymRightarrow\ nat\ list\ \isasymRightarrow\ real\ pmf\isanewline
\ \ \isakeyword{where}\ example{\isacharunderscore}{\kern0pt}3\ n\ xs\ \isacharequal\ \isanewline
\ \ \ \ do\ \isacharbraceleft{\kern0pt}\isanewline
\ \ \ \ \ \ h\ \isasymleftarrow\ sample{\isacharunderscore}{\kern0pt}pro\ \isacharequal\isacharless\isacharless\ \isasymH{\isactrlsub{P}}\ \isadigit{4}\ n\ \isacharparenleft\isasymL\ \isacharbrackleft\isadigit{1}\isacharcomma\isacharminus\isadigit{1}\isacharbrackright\isacharparenright\isanewline
\ \ \ \ \ \ return\isacharunderscore{\kern0pt}pmf\ \isacharparenleft\isacharparenleft{\isasymSum}x\ \isasymleftarrow\ xs\isachardot\ h\ x\isacharparenright\isactrlsup{\isadigit{2}}\isacharparenright\isanewline
\ \ \ \ \isacharbraceright
\end{isabelle_cm}
\Ie, we replace function applications with monadic applications \isa{\isacharequal\isacharless\isacharless} and use the constructor \isa{\isasymH{\isactrlsub{P}}} instead of the deterministic version \isa{\isasymH}.
Note that \isa{\isacharequal\isacharless\isacharless} is just the flipped version of the more classical bind operator \isa{\isasymbind}.

This actually increases the number of coin flips, but the expected number of coin flips remains the same asymptotically.

\section{Expander graphs~\label{sec:expander}}
The next pseudorandom object is based on \emph{random walks in expander graphs}~\cite{alon1986, chung1996}.
They are complementary to hash families.
While $k$-independent hash families approximate independent sampling, in the sense that any $k$ element set form independent variables, random walks in expander graphs form a more analytic approximation of independence: An $l$-step random walk in an expander graph satisfies Chernoff-bounds similar to that of an $l$-wise independent product. What this exactly means will be explored at the end of this section.

In the example in Section~\ref{sec:combining_pros}, we will see how the combination of these two types of pseudorandom objects can lead to impressive results.
Let us first briefly review the concept of expander graphs before presenting the notion of \emph{strongly explicit expander graphs}, concluding with the construction of the pseudorandom object as random walks in them.

A \emph{family of expander graphs} is an infinite sequence $G_1, G_2,\ldots$ of undirected $d$-regular graphs\footnote{A graph is $d$-regular if all of its nodes have degree $d$.} satisfying a common expansion property (detailed below).
The size of the graphs increases monotonically but not too fast.
Hoory \etal{} require $|G_i| < |G_{i+1}| \leq |G_i|^2$~\cite[Def 2.3]{hoory2006}, which we follow here, but note that the growth condition is by no means standard.
Moreover, it is essential to note that parallel edges and self-loops are allowed. Mathematically, such a graph can be modeled as a pair of finite sets $(V, E)$, where $E$ is a subset of $V \times V \times L$ and $L$ is an arbitrary space of labels for the edges.
An alternative would be to use the more classical $E \subseteq V \times V$ but allow $E$ to be a multiset.

\begin{figure}[ht]
\centering
\includegraphics[scale=0.8]{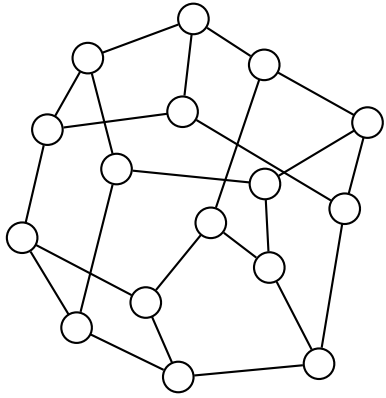}
\caption{A $3$-regular graph with expansion properties: $\lambda_a(G) \leq 0.85, \lambda_2(G) \leq 0.71, h(G) \leq 0.72$.}
\label{fig:exp_graph}
\end{figure}

For each graph, we can associate an adjacency matrix.
Given an ordering of the vertices, each cell of the adjacency matrix $A = (a_{ij})$ denotes the number of edges between the $i$-th vertex and the $j$-th vertex.
The fact that the graphs must be undirected and regular, \ie, the degree of all nodes being the same, means that the matrix is symmetric, and each row and column has the same sum corresponding to the degree of the nodes.
The normalized matrix, \ie, scaling the adjacency matrix by the inverse degree of the graph, is called the stochastic matrix.
It corresponds to the transition matrix of a random walk in the graph.
Because the matrix is symmetric the eigenvalues of both the stochastic and adjacency matrices are all on the real line.
We denote the $|G|$ eigenvalues (taking multiplicities into account) of the stochastic matrix in decreasing order by $\lambda_1,\ldots,\lambda_{|G|}$, \ie, $\lambda_1 \geq \lambda_2 \geq \cdots \geq \lambda_{|G|}$, and the second largest magnitude of the eigenvalues by $\lambda_a$, \ie, $\lambda_a := \max(|\lambda_2|, |\lambda_{|G|}|)$.\footnote{The definition is consistent with the description because $\lambda_1 = 1$ and $\lambda_{|G|} \geq -1$ for symmetric stochastic matrices.} %
Different authors use different definitions; commonly, one of the following expansion conditions is employed:
\begin{enumerate}
\item \label{en:exp_cond_1} The second largest absolute eigenvalues of the stochastic matrices of all the graphs in the family is bounded by a common value $c < 1$, \ie,  $\lambda_a(G_i) < c$ for all $i$.
\item \label{eq:exp_cond_2} The second largest eigenvalues of the stochastic matrices are bounded by a value $c < 1$, \ie, $\lambda_2(G_i) < c$ for all $i$.
\item \label{eq:exp_cond_3} There is a common strictly positive lower bound on the edge expansions $h(G_i)$, where $h(G)$ is the minimum ratio between the number of edges crossing a cut and the (smaller) size of the cut, \ie, 
$h(G) = \min_{0<|S|<|G|} \frac{E(S, G \setminus S)}{\min(|S|, |G\setminus S|)}$.  
\end{enumerate}
%In the above, the term ``second largest'' respects the multiplicities of the (magnitudes of the) eigenvalues.
%For example, if the multiplicity of the largest eigenvalue is 2, then both the largest and second largest eigenvalue is equal. %
See also Figure~\ref{fig:exp_graph} for an example graph with its expansion properties.
Property~\ref{en:exp_cond_1} of course implies Property~\ref{eq:exp_cond_2}. The converse is not true. %
The Cheeger inequality shows that Property~\ref{eq:exp_cond_2} and \ref{eq:exp_cond_3} are qualitatively equivalent:%
\[
\frac{d}{2} (1 -  \lambda_2(G)) \leq h(G) \leq d \sqrt{2(1 - \lambda_2(G))}\textrm{.}
\]
For interested readers, it is straightforward to conclude that random walks in these graphs are rapidly mixing Markov chains~\cite{chung2012, goldreich2011_2, guruswami2016}.
There is a unique stationary state, which is the uniform distribution, \ie, the probability of each node is equal in the stationary state of the chain.
The corresponding stochastic matrix has an eigenvalue $1$ with multiplicity $1$ with a corresponding eigenvector with all components identical and non-zero.
This is the Perron--Frobenius eigenvalue.
Moreover we can automatically conclude the graphs must be connected.

There are the following alternative variational definitions for $\lambda_a(G)$,  $\lambda_2(G)$, which are useful:
\begin{align}
\lambda_a(G) & := \max_{\|x\|=1, x \perp u} |\transpose{x} A x| = \max_{\|x\|=1, x \perp u} \|A x\| \label{eq:lambda_a}\\
\lambda_2(G) & := \max_{\|x\|=1, x \perp u} \transpose{x} A x \label{eq:lambda_2}
\end{align}
where $u$ is the eigenvector ($u_i := |G|^{-1/2}$) corresponding to the uniform distribution, which for expander graphs represents the unique stationary state.

The reason we are interested in expander graphs is that random walks in them approximate independent sampling, satisfying Chernoff bounds similar to ones for independent random variables.

\begin{figure}[t]
\centering
\includegraphics{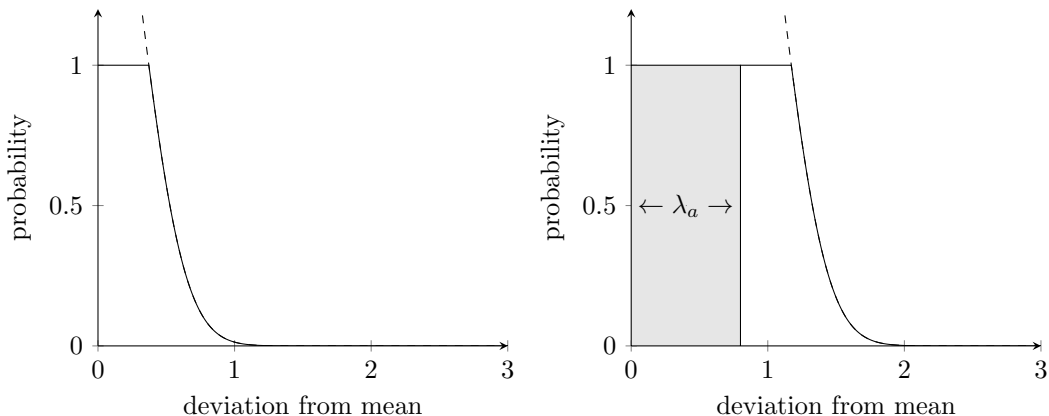}
\caption{Comparison of tail bounds for independent vs. random walk sampling (for $l=5$).}
\label{fig:tail_bound}
\end{figure}

For example, let us assume a $\mu$-fraction of the vertices of an expander graph with $n$ vertices are marked by $1$, while the other vertices are marked by $0$, \ie, let $f : G \rightarrow \{0,1\}$ such that $\mu = \frac{1}{n}|\{x\in G \mid h(x) = 1\}|$.
If we choose $l$ vertices independently, then the number of vertices hitting the $\mu$-fraction will be concentrated around $\mu n$, with a sharp tail bound satisfying:
\begin{equation}
\prob_{v \in G^l}\left( \left| \frac{1}{l} \sum_{i=1}^{l} f(v_i) - \mu \right| \geq c \right) \leq 2 \exp(- l c^2 ) \quad \textrm{for}\quad c \geq 0\textrm{.}
\end{equation}
Note that this follows from the classic Hoeffding inequality.
Instead of choosing the vertices independently, we can start at a random vertex and perform a random walk, where each edge of the walk is chosen uniformly from those that are incident to the current vertex.
As alluded to above, this will satisfy a similar Chernoff bound:
\begin{equation}
\label{eq:expander_walk_tail_bound}
\prob_{v \in \mathrm{Walk}(G,l)}\left( \left| \frac{1}{l} \sum_{i=1}^{l} f(v_i) - \mu \right| \geq (c +  \lambda_a(G)) \right) \leq 2 \exp(- l c^2 ) \quad \textrm{for}\quad c \geq 0\textrm{.}
\end{equation}
The key advantage of random walks in expander graphs, compared to choosing vertices randomly, is that the number of coin flips necessary to select such a walk is only $\bigo(\ln |G| + l \ln d)$---as opposed to $\bigo(l \ln |G|)$, \ie, we can reduce the number of coin flips, paying with the additional additive term $\lambda_a(G)$ in Eq.~\ref{eq:expander_walk_tail_bound}. (See also Figure~\ref{fig:tail_bound}.)
\subsection{Formalization}

Because of the many possible definitions, the formalization~\cite{Expander_Graphs-AFP} is as flexible as possible.
It introduces regular graphs as a specialization of general graphs, introduced in the AFP-library for Isabelle by Noschinski~\cite{Graph_Theory-AFP, noschinski2014}.
For such graphs all three types of expansion conditions are introduced.
As well as theorems such as the Cheeger inequality, Eq.~\ref{eq:lambda_a} or \ref{eq:lambda_2}, and the previously mentioned Chernoff bound for random walks (Eq.~\ref{eq:expander_walk_tail_bound}.)

The proofs rely heavily on matrix algebra.
Note that Isabelle has two major representations of matrices and vectors with different approaches.

Thiemann \etal~\cite{thiemann2016, Jordan_Normal_Form-AFP} developed a tuple-based representation, where vectors and matrices are functions paired with their dimension.
This approach is especially useful when performing induction over the dimension of the vector or matrix, \eg, verifying an algorithm that computes the Jordan normal form.
The main drawback is that they require explicit proofs of dimensional consistency.
For example to apply associativity $A (B C) = (A B) C$, we need to check that the row count of $A$ equals the column count of $B$ and that the row count of $B$ matches the column count of $C$.
This has a detrimental influence on proof automation.
 
The alternative representation---the one I chose---is part of the HOL-Library and relies on numeral types.
Vectors are represented as functions from a finite type with a size corresponding to the dimension of the vector space to the real/complex numbers.
This representation has the advantage of enabling type-level verification of dimensional consistency, which works seamlessly.
Indeed, it often helps avoid type annotations as they can be inferred automatically.
The approach was developed by Harrison~\cite{harrison2005}.

To be able to use the type-based representation, it was necessary to rely on the types-to-set mechanism~\cite{kuncar2016,kuncar2016_2}.
This works by introducing an assumption that a type exists, whose size matches the vertex set of the graph.
With the assumption it is possible to define the stochastic matrix, and perform algebraic arguments.
The assumption for the existence of the type can then be discharged for theorems that do not refer to it.
This means that the non-algebraic results about expander graphs are available even without such an assumption.
A trick I employed is to introduce $\lambda_a(G), \lambda_2(G)$ using the variational definitions (Eq.~\ref{eq:lambda_a} and \ref{eq:lambda_2}), \ie, without referring to the stochastic matrix, and show they are equivalent to the spectral definition, \ie, with respect to the multiset of eigenvalues. This allows the expression of all the relevant theorems, without explicit reference to the stochastic matrix, and hence the type pre-condition could be discharged for them.

\subsection{Diagonalization}
We will not go into all of the proofs. Nevertheless, the following is an interesting proof technique~\cite[Expander{\textunderscore}Graphs{\textunderscore}Eigenvalues]{Expander_Graphs-AFP} that highlights some of the fine issues when formalizing textbook proofs. Let us recall Eq.~\ref{eq:lambda_a}:
\begin{equation}
\label{eq:lambda_a_2}
\lambda_a(G) = \max_{\|x\|=1, x \perp u} |\transpose{x} A x|
\end{equation}
where $A$ is the stochastic matrix of a regular graph $G$, and $u$ is the vector whose components are all $|G|^{-1/2}$.

Note that this is true, whether or not $G$ is connected.
Even though expander graphs are connected, it is useful to establish Eq.~\ref{eq:lambda_a_2} unconditionally, because in the formalization it is used as an introduction rule to establish that a graph is an expander by proving the expansion condition.
This then spares us from having to separately verify the connectedness of the graph, since it follows implicitly, if we can establish $\lambda_a(G) < 1$ (or $\lambda_2(G) < 1$).

The easiest way to see that Eq.~\ref{eq:lambda_a_2} is true is to consider the diagonalization of $A$, \ie, there exists a unitary matrix $U$ \suchthat{} $A = \transpose{U} D U$ where $D$ is a diagonal matrix consisting of the eigenvalues (with their respective multiplicities) of $A$.
The rows of $U$ consist of eigenvectors of $A$.
More precisely, the $i$-th row is the eigenvector, corresponding to the $i$-th eigenvalue in the diagonal matrix $D$.
This means the right-hand side of Eq.~\ref{eq:lambda_a_2} equals $ \max_{\|y\|=1, y \perp Uu} |\transpose{y} D y|$.
One of the diagonal elements of $D$ (\eg, $d_{ii}$) must be a $1$ for which the corresponding row in $U$ should be the stationary eigenvector $u$.
And $Uu$ will be the $i$-th basis vector. From this, we can derive that the right-hand side is indeed $\lambda_a(G)$, the second largest eigenvalue of A.

Unfortunately, that argument fails, in general, if the multiplicity of the eigenvalue $1$ is larger than $1$.
However, there always exists a decomposition of the above form, where one of the rows of $U$ is the eigenvector $u$, but that is not something we can immediately derive from the classical decomposition theorem.

The best strategy to solve this problem is to use simultaneous diagonalization.
For that, observe that $A$ commutes with the matrix $J$, whose entries are all $|G|^{-1}$.
This follows from the regularity property.
On the other hand, for $J$, the multiplicity of the eigenvalue $1$ is $1$.
The only other eigenvalue of $J$ is $0$ with a multiplicity of $|G|-1$.
Now let us obtain a simultaneous diagonalization of $A$ and $J$ (this is possible because both are symmetric, real, and they commute with each other~\cite[Th.~1.3.12, Th.~2.5.3]{horn2013})\footnote{The result has been formalized by Echenim~\cite{Commuting_Hermitian-AFP}.}:
\begin{align*}
A & = \transpose{U} D U \\  J & = \transpose{U} E U
\end{align*}
Now, there is an index $i$ \suchthat{} $e_{ii} = 1$, corresponding to the $u$ in $U$, because of the decomposition of $J$.
Thus, we obtain a sequence of eigenvectors of $A$, which necessarily contains the eigenvector $u$.

It should be noted that the same result could be obtained by introducing a correction to the diagonalization of $A$ since the span of the eigenspace for the eigenvalue $1$ will contain the eigenvector $u$.
However, the approach above is much more elegant. 

\subsection{Strongly explicit expander graphs}
The class of expander graphs that are needed for derandomization are strongly explicit expander graphs.
For such a graph, it is possible to sample a random vertex uniformly and compute the edges incident to a given vertex algorithmically, \ie, it is possible to sample a random walk without having to represent the graph in memory.
This makes them very useful for the application in space-constrained settings.
This is in contrast to randomized constructions of expander graphs~\cite{alon1986,mckay1981}, which is much easier but would require the representation of the graph in memory.
%\footnote{Edmonds and Paulson~\cite{edmonds2024} have done foundational work on the formalization of probability theoretic existence proofs for (hyper-)graphs.}

The first strongly explicit construction was by Margulis~\cite{margulis1973}, who could show for a family of graphs that they have a spectral bound strictly smaller than $1$, but the bound itself was unknown.
A few years later, Gabber and Galil~\cite{gabber1981} managed to show, for a slightly different family of graphs, a concrete spectral bound: $\lambda_a(G_n) \leq \frac{5}{8}\sqrt{2} \approx 0.884$.

The description of their family of graphs---now usually called the Margulis--Gaber--Galil family---is pretty elementary. For every $n > 0$, the graph $G_n$ has $n^2$ vertices, which we label with the elements of $\mathbb{Z}/n\mathbb{Z} \times \mathbb{Z}/n\mathbb{Z}$. Then each vertex $(x,y) \in G_n$ is connected to the vertices:
\[
   (x+1,y), (x-1,y) (x,y+1), (x,y-1), (x,y+x), (x,y-x), (x+y,y), (x-y,y)
\]
The elementary description is, however, quite deceiving because the derivation of the spectral bound is far from it.
There were multiple efforts towards reducing the complexity of the proof.
Indeed, the initial proof by Gabber and Galil was improved by Jimbo and Marouka~\cite{jimbo1985} in 1985.
Ravi Boppana further improved it in his lectures.
Hoory \etal~\cite[\S 8]{hoory2006} present a version of his proof.
But even for that last version, the authors admit that it is ``subtle and mysterious.''
It heavily relies on the discrete Fourier transform on the torus $\mathbb{Z}/n\mathbb{Z} \times \mathbb{Z}/n\mathbb{Z}$, followed by a number of careful decoupling steps, a partial order, many case splits, and symmetry arguments.

After that, there was more research.
For example, Lee~\cite{lee2024} shows a weaker spectral bound for the same family but with a less complex proof.

The formalized proof~\cite[Expander{\textunderscore}Graphs{\textunderscore}MGG]{Expander_Graphs-AFP} obtains the best-known spectral bound $\frac{5}{8}\sqrt{2}$ and follows the previously mentioned proof by Hoory \etal{}
(although I could improve on their proof slightly by using one more symmetry argument to reduce the number of case splits).

After Margulis's discovery, many more constructions have been found~\cite{alon2007, marcus2015}.
An interesting solution was discovered by Reingold \etal~\cite{reingold2000} called the zig-zag construction, which works by starting with small fixed expander graphs and performing various graph operations to construct a family of larger expander graphs.
This would be easy to formalize, although the Margulis construction is much more elegant.
Moreover, the fact that the vertex counts of successive Margulis graphs do not grow by more than a factor of two is helpful, as we will see later in this section.

Other constructions rely on the Cayley graph of non-abelian groups~\cite{kassabov2006, lubotzky2011}.
The proofs heavily rely on representation theory, which would require much more groundwork in Isabelle/HOL.
In particular, constructions of Ramanujan graphs~\cite{lubotzky1988}, which are expander graphs with the best-possible spectral bound (with respect to their degree.), belong to this group.
I believe the formalization of such a construction is a significant open problem.

\subsection{Graph operations}
The Margulis--Gabber--Galil family consists of graphs for all vertex counts that are square. Using a method described by Murtagh \etal~\cite{murtagh2019}, it is possible to obtain expander graphs for every vertex count.
This works by vertex identification, specifically given an expander graph $G$ with degree $d$, a spectral bound $\lambda_a$ and $n$ vertices $v_0,\ldots,v_{n-1}$, we identify distinct vertices (\eg, vertex $v_0$ with $v_m$, $v_1$ with $v_{m+1}$ etc. and $v_{n-m-1}$ with $v_{n-1}$) for some $m$ \suchthat{} $\frac{n}{2} \leq m \leq n$.
Note that edges are preserved, even if this introduces parallel edges.
If there is an edge between identified vertices, the edge becomes a self-loop.
To preserve regularity, we introduce $d$ self-loops for vertices that have not been paired (\ie{} for $v_{n-m},\ldots,v_{m-1}$).
This means the resulting contracted graph will have degree $2d$.

Then, the resulting graph will have a spectral bound of at most $\frac{\lambda_a(G)+1}{2}$.

While the idea is due to Murtagh \etal, they only show that contracting a graph with a spectral bound of at most $\frac{1}{4}$ results in a graph with a spectral bound of at most $\frac{1}{2}$.
Note that the result stated above, which is part of the formalization~\cite[Expander{\textunderscore}Graphs{\textunderscore}Strongly{\textunderscore}Explicit]{Expander_Graphs-AFP}, is a generalization.

Let us go through the proof, which can be verified using a generalization of Eq.~\ref{eq:lambda_a}.
\begin{equation}
\label{eq:lambda_a_gen}
	|\transpose{x} A x| \leq \lambda_a(G) \| x \|^2 + (1 - \lambda_a(G)) (\transpose{x} u)^2 \quad \textrm{for \emph{all}}\quad x\textrm{.}
\end{equation}
where $u$ is defined as in Eq.~\ref{eq:lambda_a}.

Let $x \in \mathbb{R}^m$, $x  \neq 0$ and $x \perp u'$ and let $A$ denote the stochastic matrix of $G$ and $A'$ be the stochastic matrix of the contracted graph $G'$. 
Similarly $u \in \mathbb{R}^n, u_i = n^{-1/2}$ and $u' \in \mathbb{R}^m, u'_i = m^{-1/2}$. 
Moreover let $y \in \mathbb{R}^n$ \suchthat{} $y_i := x_{i \omod m}$ and $\beta = \left|\sum_{i=n-m}^{m-1} x_i^2\right|$. Then:

\begingroup
\allowdisplaybreaks
\begin{eqnarray*}
2 |\transpose{x} A' x| & = & \left|\sum_{i,j=0}^{n-1} x_{i \omod m} a_{ij} x_{j \omod m} + \sum_{i=n-m}^{m-1} x_i^2\right| \leq |\transpose{y} A y| + \beta \\
& \stackunder{$\leq$}{\footnotesize Eq.~\ref{eq:lambda_a_gen}} & \lambda_a(G) \|y\|^2 + (1-\lambda_a(G)) (\transpose{y} u)^2 + \beta \\
& = &  \lambda_a(G) (\|y\|^2 + \beta) + (1-\lambda_a(G)) ((\transpose{y} u)^2 + \beta) \\
& = & 2 \lambda_a(G) \|x\|^2 + (1-\lambda_a(G)) \left(\left(\sum_{i=0}^{n-1} x_{i \omod m} n^{-1/2} \right)^2 + \beta\right) \\
& \stackunder{$=$}{\footnotesize $x \perp u'$} & 2 \lambda_a(G) \|x\|^2 + (1-\lambda_a(G)) \left(\left(\sum_{i=m}^{n-1} x_{i \omod m} n^{-1/2} \right)^2 + \beta\right) \\
& \stackunder{$\leq$}{\footnotesize Cauchy--Schwarz} &  2 \lambda_a(G) \|x\|^2 + (1-\lambda_a(G)) \left(\left(\sum_{i=m}^{n-1} x_{i \omod m}^2\right)\left(\sum_{i=m}^{n-1} \frac{1}{n}\right) + \beta\right) \\ % && \textrm{By Cauchy-Schwarz.}
& = & 2 \lambda_a(G) \|x\|^2 + (1-\lambda_a(G)) \left(\frac{n-m}{n} \left(\sum_{i=0}^{n-m-1} x_i^2\right) + \beta\right) \\
& \stackunder{$\leq$}{\footnotesize $(n-m)/n \leq 1$} & 2 \lambda_a(G) \|x\|^2 + (1-\lambda_a(G)) \|x\|^2 \textrm{.\qed}
\end{eqnarray*}
\endgroup

Using the Margulis construction and the fact that there is a square between any $n$ and $2n$ (inclusive), it follows that there is a strongly explicit expander graph with spectral bound $(5 \sqrt{2} + 8)/16$ and degree $16$ for every vertex count. 

Another graph operation is the power operation, which can be used to improve the spectral bound~\cite[Expander{\textunderscore}Graphs{\textunderscore}Power{\textunderscore}Construction]{Expander_Graphs-AFP}. Let $G$ be a graph, then the $k$-th power of $G$ is a graph---denoted as $G^k$---consisting of the same set of vertices, where the number of edges between two vertices $v$ and $w$ in $G^k$ is the number of paths of length $k$ between $v$ and $w$ in $G$. Note that algebraically the stochastic matrix of $G^k$ is $A^k$ if $A$ is the stochastic matrix of $G$.
It is easy to verify that $\lambda_a(G^k) \leq \lambda_a(G)^k$.

The previous two observations, in combination, imply that there exists a strongly explicit expander graph for every vertex count and any spectral bound $\lambda_a>0$ with a degree satisfying: $\ln d \in \bigo( \ln (\lambda_a^{-1}))$ as $\lambda_a \rightarrow 0$~\cite[Expander{\textunderscore}Graphs{\textunderscore}Strongly{\textunderscore}Explicit]{Expander_Graphs-AFP}.
At this point, we are finally at a stage where we can introduce the second pseudorandom object, denoted by \isa{\isasymE\ l\ \isasymLambda\ P}.
It first selects the strongly explicit expander graph constructed above with a number of vertices $n$, matching the size of the seed space of $P$ and a spectral bound of at most \isa{\isasymLambda}.
Then, it samples random walks in that graph of length $l$.
Note that because of the regularity of the graph, there are exactly $n d^{l-1}$ random walks, and given a number between $\{0,\ldots,nd^{l-1}-1\}$, we can output the vertices representing that path algorithmically.
Here, it is important that multiplicities are taken into account, which is also the reason some vertex tuples may have a multiplicity higher than $1$.
   
Note that, for the users of the library, it is possible to ignore the complex internal structure of the pseudorandom object. Let us consider the classic Hoeffding bound for independent $\{0,1\}$-valued random variables:
\begin{isabelle_cm}
\isacommand{theorem}\ classic{\isacharunderscore}chernoff{\isacharunderscore}bound\isacharcolon\isanewline
\ \ \isacommand{assumes}\ AE\ x\ in\ measure{\isacharunderscore}pmf\ p\isachardot\ f\ x\ \isasymin\ \isacharbraceleft\isadigit{0}\isacharcomma\isadigit{1}\isacharbraceright\isanewline
\ \ \isacommand{defines}\ \isasymmu\ =\ expectation\ p\ f\isanewline
\ \ \isacommand{shows} \isasymP{\isacharparenleft}w\ in\ prod{\isacharunderscore}pmf\isacharbraceleft\isadigit{0}\isachardot\isachardot{\isacharless}l\isacharbraceright\ \isacharparenleft\isasymlambda{\isacharunderscore}\isachardot\ p\isacharparenright\isachardot\ \isacharbar{\isasymSum}j{\isacharless}l\isachardot\ f{\isacharparenleft}w\ j\isacharparenright{\isacharminus}l\isacharasterisk\isasymmu\isacharbar{\isasymge}l{\isacharasterisk}c\isacharparenright\isasymle\isadigit{2}{\isacharasterisk}exp\isacharparenleft{\isacharminus}n{\isacharasterisk}c\isacharcircum\isadigit{2}\isacharparenright
\end{isabelle_cm}
If we instead wanted to choose expander random walks, we can substitute the sampling space and use the Chernoff bound for these~\cite[Pseudorandom{\textunderscore}Objects{\textunderscore}Expander{\textunderscore}Walks]{Expander_Graphs-AFP}:

\begin{isabelle_cm}
\isacommand{theorem}\ expander{\isacharunderscore}chernoff{\isacharunderscore}bound\isacharcolon\isanewline
\ \ \isacommand{assumes}\ AE\ x\ in\ sample{\isacharunderscore}pro\ P\isachardot\ f\ x\ \isasymin\ \isacharbraceleft\isadigit{0}\isacharcomma\isadigit{1}\isacharbraceright\isanewline
\ \ \isacommand{defines}\ \isasymmu\ =\ expectation\ {\isacharparenleft}sample{\isacharunderscore}pro\ P\isacharparenright\ f\isanewline
\ \ \isacommand{shows} \isasymP{\isacharparenleft}w\ in\ sample{\isacharunderscore}pro\ \isacharparenleft\isasymE\ l\ \isasymLambda\ P\isacharparenright\isachardot\ \isacharbar{\isasymSum}j{\isacharless}l\isachardot\ f{\isacharparenleft}w\ j\isacharparenright{\isacharminus}l\isacharasterisk\isasymmu\isacharbar{\isasymge}l{\isacharasterisk}{\isacharparenleft}c\isacharplus\isasymLambda\isacharparenright\isacharparenright\isasymle\isadigit{2}{\isacharasterisk}exp\isacharparenleft{\isacharminus}n{\isacharasterisk}c\isacharcircum\isadigit{2}\isacharparenright
\end{isabelle_cm}

Like before, the theorem is expressed with respect to the probability space of the chained pseudorandom object: \isa{sample{\isacharunderscore}pro\ P}, which allows the derivation of results with respect to more complex combined pseudorandom objects seamlessly. As mentioned before Section~\ref{sec:combining_pros} will demonstrate this on a concrete example.  

\subsection{Efficiency}
The pseudorandom objects presented are implemented in an effective and efficient way.
This means that the algorithms can be exported into executable efficient code~\cite{haftmann2010} in various languages such as Haskell, Scala or SML.
For example, the field inversion operation is implemented using Euclid's algorithm.
Large powers are computed using the exponentiation by squaring technique.

Here, it is essential to note that, for randomized algorithms, this does not work directly because they are represented in Isabelle using the Giry monad.
However, there is a framework introduced by Eberl and myself~\cite{Executable_Randomized_Algorithms-AFP}, which enables exporting code for randomized algorithms.
This works by externalizing the source of randomness using a lazy stream of coin flips and establishing a Scott-continuous monad-morphism between the Giry monad representation and the representation as a randomness consumer.
The resulting code can then be executed in the target language using any source of randomness (for example, a hardware random number generator).

\section{Combining pseudorandom objects~\label{sec:combining_pros}}
This section presents a concrete example where a combination of multiple pseudorandom objects are used to improve an algorithm's space complexity.
To achieve that, we are going to start with \isa{example{\isacharunderscore}2} from the introduction and improve its accuracy (relative error with respect to the correct value) and its probability of success, \ie,
the probability with which the algorithm returns a result with the desired accuracy. To make the presentation accessible, we will do this step-by-step and also briefly point to the 
main points of the correctness proofs. The examples are all also available in the AFP~\cite[Tutorial{\textunderscore}Pseudorandom{\textunderscore}Objects]{Frequency_Moments-AFP}. 
Let us recall:
\begin{isabelle_cm}
\isacommand{fun}\ example{\isacharunderscore}{\kern0pt}2\ \isacharcolon\isacharcolon\ nat\ \isasymRightarrow\ nat\ list\ \isasymRightarrow\ real\ pmf\isanewline
\ \ \isakeyword{where}\ example{\isacharunderscore}{\kern0pt}2\ n\ xs\ \isacharequal\ \isanewline
\ \ \ \ do\ \isacharbraceleft{\kern0pt}\isanewline
\ \ \ \ \ \ h\ \isasymleftarrow\ sample{\isacharunderscore}{\kern0pt}pro\ \isacharparenleft\isasymH\ \isadigit{4}\ n\ \isacharparenleft\isasymL\ \isacharbrackleft\isadigit{1}\isacharcomma\isacharminus\isadigit{1}\isacharbrackright\isacharparenright\isacharparenright\isanewline
\ \ \ \ \ \ return\isacharunderscore{\kern0pt}pmf\ \isacharparenleft\isacharparenleft{\isasymSum}x\ \isasymleftarrow\ xs\isachardot\ h\ x\isacharparenright\isactrlsup {\isadigit{2}}\isacharparenright\isanewline
\ \ \ \ \isacharbraceright
\end{isabelle_cm}
We had already established that the result has the expectation $F_2$ and a variance of at most $2F_2^2$.
A standard technique to improve the accuracy of such an algorithm is to run it in parallel multiple times and obtain the average.
\begin{isabelle_cm}
\isacommand{fun}\ example{\isacharunderscore}{\kern0pt}4\ \isacharcolon\isacharcolon\ nat\ \isasymRightarrow\ nat\ list\ \isasymRightarrow\ real\ pmf\isanewline
\ \ \isakeyword{where}\ example{\isacharunderscore}{\kern0pt}4\ n\ xs\ \isacharequal\ \isanewline
\ \ \ \ do\ \isacharbraceleft{\kern0pt}\isanewline
\ \ \ \ \ \ let\ s\ \isacharequal\ \isasymlceil8\isacharslash\isasymepsilon\isacharcircum\isadigit{2}\isasymrceil\isacharsemicolon\isanewline
\ \ \ \ \ \ h\ \isasymleftarrow\ prod{\isacharunderscore}pmf \isacharbraceleft\isadigit{0}\isachardot\isachardot{\isacharless}s\isacharbraceright\ \isacharparenleft\isasymlambda\isacharunderscore\isachardot\ sample{\isacharunderscore}{\kern0pt}pro\ \isacharparenleft\isasymH\ \isadigit{4}\ n\ \isacharparenleft\isasymL\ \isacharbrackleft\isadigit{1}\isacharcomma\isacharminus\isadigit{1}\isacharbrackright\isacharparenright\isacharparenright\isacharparenright\isacharsemicolon\isanewline
\ \ \ \ \ \ return\isacharunderscore{\kern0pt}pmf\ \isacharparenleft\isacharparenleft{\isasymSum}j{\isacharless}s\isachardot\ \isacharparenleft{\isasymSum}x\ \isasymleftarrow\ xs\isachardot\ h\ j\ x\isacharparenright\isactrlsup {\isadigit{2}}\isacharparenright{\isacharslash}s\isacharparenright\isanewline
\ \ \ \ \isacharbraceright
\end{isabelle_cm}
It is possible to show that the above algorithm returns an estimate of $F_2$ with relative error $\varepsilon$ with probability at least $\frac{3}{4}$, or equivalently the probability that the result does not have the desired accuracy is at most $\frac{1}{4}$.
%(Note in general for such probability theoretic results, it is more convenient to verify an upper bound for the failure probability, than to establish a lower bound for the success probability.
%This is mostly due to the fact that establishing increasing inequality chains are more natural, as well as the fact that many already verified results such as the Chebyshev inequality or Hoeffding inequality are expressed in a way that more naturally fit with the upper-bound-approach.)
\begin{isabelle_cm}
\isacommand{lemma}\ example{\isacharunderscore}4{\isacharunderscore}correct\isacharcolon\isanewline
\ \ \isacommand{assumes}\ set\ xs\ \isasymsubseteq\ \isacharbraceleft\isadigit{0}\isachardot\isachardot{\isacharless}n\isacharbraceright\isanewline
\ \ \isacommand{shows}\ \isasymP\isacharparenleft\isasymomega\ in\ example{\isacharunderscore}4\isachardot\ \isacharbar\isasymomega\ \isacharminus\ F2\ xs\isacharbar\ \isachargreater\ \isasymepsilon\ \isacharasterisk\ F2\ xs\isacharparenright\ \isasymle\ 1{\isacharslash}4
\end{isabelle_cm}
Usually the first step of the verification is to express the result in terms of random variables over the probability space of the pseudorandom object(s):
\begin{isabelle_cm}
\ \ \isacommand{define}\ s\ \isacommand{where}\ s\ \isacharequal\ \isasymlceil8\isacharslash\isasymepsilon\isacharcircum\isadigit{2}\isasymrceil\isanewline
\ \ \isacommand{define}\ R\ \isacommand{where}\ R\ h\ \isacharequal\ \isacharparenleft{\isasymSum}j{\isacharless}s\isachardot\ \isacharparenleft{\isasymSum}x\ \isasymleftarrow\ xs\isachardot\ h\ x\isacharparenright\isactrlsup {\isadigit{2}}\isacharparenright{\isacharslash}s\isanewline
\isanewline
\ \ \isacommand{let}\ {\isacharquery}q\ \isacharequal\ prod{\isacharunderscore}pmf \isacharbraceleft\isadigit{0}\isachardot\isachardot{\isacharless}s\isacharbraceright\ \isacharparenleft\isasymlambda\isacharunderscore\isachardot\ sample{\isacharunderscore}{\kern0pt}pro\ \isacharparenleft\isasymH\ \isadigit{4}\ n\ \isacharparenleft\isasymL\ \isacharbrackleft\isadigit{1}\isacharcomma\isacharminus\isadigit{1}\isacharbrackright\isacharparenright\isacharparenright\isacharparenright
\end{isabelle_cm}
With these definition it is now straightforward to check that the left-hand-side of the lemma equals:
\begin{isabelle_cm}
\ \ measure\ {\isacharquery}q\ {\isacharbraceleft}h\isachardot\ {\isacharbar}R\ h\ \isacharminus\ F2\ xs{\isacharbar}\ \isachargreater\ \isasymepsilon\ \isacharasterisk\ F2\ xs\isacharbraceright
\end{isabelle_cm}
which can be shown to be at most $\frac{1}{4}$ using Chebyshev's inequality and
the facts that:
\begin{itemize} 
  \item The expectation of \isa{R} is \isa{F2\ xs}.
  \item The variance is at most \isa{\isasymepsilon{\isacharcircum}2\ \isacharasterisk\ F2\ xs{\isacharcircum}2{\isacharslash}4}.
\end{itemize}
Both results follow using the previously mentioned result about the expectation and variance of \isa{example{\isacharunderscore}2}.
The first one follows directly using linearity of expectations. The second condition requires the application of Bienaym\'e's identity, \ie, the variance of the sum of \emph{pairwise independent} random variables equals the sum of the variances of the random variables:
\begin{isabelle_cm}
\isacommand{lemma}\ bienaymes{\isacharunderscore}identity{\isacharunderscore}pairwise{\isacharunderscore}indep{\isacharunderscore}2\isacharcolon\footnotemark\isanewline
\ \ \isacommand{fixes}\ f\ \isacharcolon\isacharcolon{\isacharprime}b\ \isasymRightarrow\ {\isacharprime}a\ \isasymRightarrow\ real\isanewline
\ \ \isacommand{assumes}\ finite\ I\isanewline
\ \ \isacommand{assumes}\ {\isasymAnd}i\isachardot\ i\ \isasymin\ I\ \isasymLongrightarrow\ f\ i\ \isasymin\ borel{\isacharunderscore}measurable\ M\isanewline
\ \ \isacommand{assumes}\ {\isasymAnd}i\isachardot\ i\ \isasymin\ I\ \isasymLongrightarrow\ integrable\ M\ \isacharparenleft\isasymlambda\isasymomega\isachardot\ f\ i\ \isasymomega{\isacharcircum}2\isacharparenright\isanewline
\ \ \isacommand{assumes}\ {\isasymAnd}J\isachardot\ J\ \isasymsubseteq\ I\ \isasymLongrightarrow\ card\ J\ \isacharequal\ 2\ \isasymLongrightarrow\ indep{\isacharunderscore}vars\ \isacharparenleft\isasymlambda\isacharunderscore\isachardot\ borel\isacharparenright\ f\ J\isanewline
\ \ \isacommand{shows}\ variance\ \isacharparenleft\isasymlambda\isasymomega\isachardot\ \isacharparenleft{\isasymSum}i\ \isasymin\ I\isachardot\ f\ i\ \isasymomega\isacharparenright\isacharparenright\ \isacharequal\ \isacharparenleft{\isasymSum}i\ \isasymin\ I\isachardot\ variance\ {\isacharparenleft}f\ i\isacharparenright\isacharparenright
\end{isabelle_cm}
\footnotetext{\cite[Bienaymes{\textunderscore}Identity]{Concentration_Inequalities-AFP}}
The above solution requires $\bigo( \varepsilon^{-2} (\ln n + \ln m) )$ bits of space.
However, it is possible to improve on that considerably because Bienaym\'{e}'s identity only requires pairwise independence,
something we already know how to handle. Instead of choosing the seeds of the hash functions independently, it is enough to choose them pairwise independently, which can again be done using hash families.
The following is the improved version, where we replaced the product construction with the pseudorandom object for $2$-independent hash families:
\begin{isabelle_cm}
\isacommand{fun}\ example{\isacharunderscore}{\kern0pt}5\ \isacharcolon\isacharcolon\ nat\ \isasymRightarrow\ nat\ list\ \isasymRightarrow\ real\ pmf\isanewline
\ \ \isakeyword{where}\ example{\isacharunderscore}{\kern0pt}5\ n\ xs\ \isacharequal\ \isanewline
\ \ \ \ do\ \isacharbraceleft{\kern0pt}\isanewline
\ \ \ \ \ \ let\ s\ \isacharequal\ \isasymlceil8\isacharslash\isasymepsilon\isacharcircum\isadigit{2}\isasymrceil\isacharsemicolon\isanewline
\ \ \ \ \ \ h\ \isasymleftarrow\ sample{\isacharunderscore}{\kern0pt}pro\ \isacharparenleft\isasymH\ \isadigit{2}\ s\ \isacharparenleft\isasymH\ \isadigit{4}\ n\ \isacharparenleft\isasymL\ \isacharbrackleft\isadigit{1}\isacharcomma\isacharminus\isadigit{1}\isacharbrackright\isacharparenright\isacharparenright\isacharparenright\isacharsemicolon\isanewline
\ \ \ \ \ \ return\isacharunderscore{\kern0pt}pmf\ \isacharparenleft\isacharparenleft{\isasymSum}j{\isacharless}s\isachardot\ \isacharparenleft{\isasymSum}x\ \isasymleftarrow\ xs\isachardot\ h\ j\ x\isacharparenright\isactrlsup {\isadigit{2}}\isacharparenright{\isacharslash}s\isacharparenright\isanewline
\ \ \ \ \isacharbraceright
\end{isabelle_cm}
The correctness proof for \isa{example{\isacharunderscore}5} and \isa{example{\isacharunderscore}4} are, of course, very similar.
The additional lemmas~\cite[Pseudorandom{\textunderscore}Objects{\textunderscore}Hash{\textunderscore}Families]{Universal_Hash_Families-AFP} needed in the proof include the fact that the outer hash family is pairwise independent:
\begin{isabelle_cm}
\isacommand{lemma}\ hash{\isacharunderscore}pro{\isacharunderscore}pmf{\isacharunderscore}k{\isacharunderscore}indep\isacharcolon\isanewline
\ \ \isacommand{assumes}\ is{\isacharunderscore}prime{\isacharunderscore}power\ {\isacharparenleft}pro{\isacharunderscore}size\ S\isacharparenright\isanewline
\ \ \isacommand{shows}\ prob{\isacharunderscore}space{\isachardot}k{\isacharunderscore}wise{\isacharunderscore}indep{\isacharunderscore}vars\ {\isacharparenleft}sample{\isacharunderscore}pro\ \isacharparenleft\isasymH\ k\ n\ S\isacharparenright\isacharparenright\ k\ \isacharparenleft\isasymlambda\isacharunderscore\isachardot\ discrete\isacharparenright\ \isacharparenleft{\isasymlambda}i\ \isasymomega\isachardot\ \isasymomega\ i\isacharparenright\ \isacharbraceleft\isachardot\isachardot{\isacharless}n\isacharbraceright
\end{isabelle_cm}
and the fact that mapping to a specific component is like sampling directly from the inner pseudorandom object.
\begin{isabelle_cm}
\isacommand{lemma}\ hash{\isacharunderscore}pro{\isacharunderscore}pmf{\isacharunderscore}component\isacharcolon\isanewline
\ \ \isacommand{assumes}\ is{\isacharunderscore}prime{\isacharunderscore}power\ {\isacharparenleft}pro{\isacharunderscore}size\ S\isacharparenright\isanewline
\ \ \isacommand{assumes}\ i\ \isacharless\ n\isanewline
\ \ \isacommand{assumes}\ k\ \isachargreater\ 0\isanewline
\ \ \isacommand{shows}\ map{\isacharunderscore}pmf\ \isacharparenleft{\isasymlambda}f\isachardot\ f\ i\isacharparenright\ {\isacharparenleft}sample{\isacharunderscore}pro\ \isacharparenleft\isasymH\ k\ n\ S\isacharparenright\isacharparenright\ \isacharequal\ sample{\isacharunderscore}pro\ S
\end{isabelle_cm}
The latter theorem illustrates a common approach when working with composed pseudorandom objects.
First a result is established for the inner pseudorandom object, then a theorem like the above can be used to lift it to a statement for the combined object. 
Indeed, the theorems about pseudorandom objects are specifically designed in a way that makes this possible. 
 
Note that the resulting algorithm now requires only $\bigo( \ln n + \varepsilon^{-2} \ln m )$ bits.
This algorithm has a failure probability of $\frac{1}{4}$. In the next step, we would like to improve the failure probability.

A straightforward method would be to increase the parameter $s$ (the number of repetitions), \ie, to obtain a mean of a larger sample.
Indeed, by setting $s = \lceil 2 \varepsilon^{-2} \delta^{-1} \rceil$, the algorithm would return an $\varepsilon$-approximation with a probability of at least $(1-\delta)$. However, there is a much more efficient strategy.

Let us first observe that the median of a sequence will be an $\varepsilon$-approximation if more than half of the sequence elements are.
(Note that under the assumption, when we sort the sequence, the sequence elements that are $\varepsilon$-approximations will form a consecutive subsequence, whose length would be longer than half the length of the sequence. 
Thus, the subsequence would necessarily include the median.)
This is reflected in the following theorem~\cite{Median_Method-AFP}:
\begin{isabelle_cm}
\isacommand{lemma}\ median{\isacharunderscore}est\isacharcolon\isanewline
\ \ \isacommand{assumes}\ interval\ I\isanewline
\ \ \isacommand{assumes}\ 2{\isacharasterisk}card\ {\isacharbraceleft}k\isachardot\ k\ \isacharless\ n\ \isasymand\ f\ k\ \isasymin\ I\isacharbraceright\ \isachargreater\ n\isanewline
\ \ \isacommand{shows}\ median n f \isasymin I
\end{isabelle_cm}
In proofs the converse is more applicable:
\begin{isabelle_cm}
\isacommand{lemma}\ median{\isacharunderscore}est{\isacharunderscore}rev\isacharcolon
\ interval\ I\ \isasymLongrightarrow\ median n f \isasymnotin I\ \isasymLongrightarrow\ 2{\isacharasterisk}card\ {\isacharbraceleft}k\isachardot\ k{\isacharless}n{\isasymand}f\ k\ \isasymnotin\ I\isacharbraceright{\isasymge}n
\end{isabelle_cm}
Now, what we can do is to repeat the algorithm \isa{example{\isacharunderscore}5} $\lceil 8 \ln (\delta^{-1})\rceil$ times independently and obtain the median:
\begin{isabelle_cm}
\isacommand{fun}\ example{\isacharunderscore}{\kern0pt}6\ \isacharcolon\isacharcolon\ nat\ \isasymRightarrow\ nat\ list\ \isasymRightarrow\ real\ pmf\isanewline
\ \ \isakeyword{where}\ example{\isacharunderscore}{\kern0pt}6\ n\ xs\ \isacharequal\ \isanewline
\ \ \ \ do\ \isacharbraceleft{\kern0pt}\isanewline
\ \ \ \ \ \ let\ s\ \isacharequal\ \isasymlceil8\isacharslash\isasymepsilon\isacharcircum\isadigit{2}\isasymrceil\isacharsemicolon\ let\ t\ \isacharequal\ \isasymlceil8{\isacharasterisk}ln\ {\isacharparenleft}1\isacharslash\isasymdelta\isacharparenright\isasymrceil\isacharsemicolon\isanewline
\ \ \ \ \ \ h\ \isasymleftarrow\ prod{\isacharunderscore}pmf\ \isacharbraceleft\isadigit{0}\isachardot\isachardot{\isacharless}t\isacharbraceright\ \isacharparenleft\isasymlambda\isacharunderscore\isachardot\ sample{\isacharunderscore}{\kern0pt}pro\ \isacharparenleft\isasymH\ \isadigit{2}\ s\ \isacharparenleft\isasymH\ \isadigit{4}\ n\ \isacharparenleft\isasymL\ \isacharbrackleft\isadigit{1}\isacharcomma\isacharminus\isadigit{1}\isacharbrackright\isacharparenright\isacharparenright\isacharparenright\isacharparenright\isacharsemicolon\isanewline
\ \ \ \ \ \ return\isacharunderscore{\kern0pt}pmf\ {\isacharparenleft}median\ t\ \isacharparenleft{\isasymlambda}i\isachardot\ \isacharparenleft{\isasymSum}j{\isacharless}s\isachardot\ \isacharparenleft{\isasymSum}x\ \isasymleftarrow\ xs\isachardot\ h\ i\ j\ x\isacharparenright\isactrlsup {\isadigit{2}}\isacharparenright{\isacharslash}s\isacharparenright\isacharparenright\isanewline
\ \ \ \ \isacharbraceright
\end{isabelle_cm}
Because the failure probability of \isa{example{\isacharunderscore}5} was $\frac{1}{4}$, the probability that more than half of the $t$ independent repetitions will fail (and hence the median) is at most $\exp(-\frac{t}{8}) \leq \delta$. This follows from the one-sided version of Hoeffding's inequality for $\{0,1\}$-valued independent random variables.

The above leads to an algorithm with the space usage $\bigo(\ln(\delta^{-1}) (\ln n + \varepsilon^{-2} \ln m))$.
However, even the above can be improved using the second pseudorandom object discussed in Section~\ref{sec:expander}: Random walks in expander graphs.
In Section~\ref{sec:expander}, we pointed out that they satisfy a similar Chernoff bound. Hence, we can again replace the independent sampling with a pseudorandom object. The result is the following algorithm:

\begin{isabelle_cm}
\isacommand{fun}\ example{\isacharunderscore}{\kern0pt}7\ \isacharcolon\isacharcolon\ nat\ \isasymRightarrow\ nat\ list\ \isasymRightarrow\ real\ pmf\isanewline
\ \ \isakeyword{where}\ example{\isacharunderscore}{\kern0pt}7\ n\ xs\ \isacharequal\ \isanewline
\ \ \ \ do\ \isacharbraceleft{\kern0pt}\isanewline
\ \ \ \ \ \ let\ s\ \isacharequal\ \isasymlceil8\isacharslash\isasymepsilon\isacharcircum\isadigit{2}\isasymrceil\isacharsemicolon\ let\ t\ \isacharequal\ \isasymlceil8{\isacharasterisk}ln\ {\isacharparenleft}1\isacharslash\isasymdelta\isacharparenright\isasymrceil\isacharsemicolon\isanewline
\ \ \ \ \ \ h\ \isasymleftarrow\ sample{\isacharunderscore}{\kern0pt}pro\ \isacharparenleft\isasymE\ t\ {\isacharparenleft}1{\isacharslash}8\isacharparenright\ \isacharparenleft\isasymH\ \isadigit{2}\ s\ \isacharparenleft\isasymH\ \isadigit{4}\ n\ \isacharparenleft\isasymL\ \isacharbrackleft\isadigit{1}\isacharcomma\isacharminus\isadigit{1}\isacharbrackright\isacharparenright\isacharparenright\isacharparenright\isacharparenright\isacharsemicolon\isanewline
\ \ \ \ \ \ return\isacharunderscore{\kern0pt}pmf\ {\isacharparenleft}median\ t\ \isacharparenleft{\isasymlambda}i\isachardot\ \isacharparenleft{\isasymSum}j{\isacharless}s\isachardot\ \isacharparenleft{\isasymSum}x\ \isasymleftarrow\ xs\isachardot\ h\ i\ j\ x\isacharparenright\isactrlsup {\isadigit{2}}\isacharparenright{\isacharslash}s\isacharparenright\isacharparenright\isanewline
\ \ \ \ \isacharbraceright
\end{isabelle_cm}
for which we can show:
\begin{isabelle_cm}
\isacommand{lemma}\ example{\isacharunderscore}7{\isacharunderscore}correct\isacharcolon\isanewline
\ \ \isacommand{assumes}\ set\ xs\ \isasymsubseteq\ \isacharbraceleft\isadigit{0}\isachardot\isachardot{\isacharless}n\isacharbraceright\isanewline
\ \ \isacommand{shows}\ \isasymP\isacharparenleft\isasymomega\ in\ example{\isacharunderscore}7\isachardot\ \isacharbar\isasymomega\ \isacharminus\ F2\ xs\isacharbar\ \isachargreater\ \isasymepsilon\ \isacharasterisk\ F2\ xs\isacharparenright\ \isasymle\ \isasymdelta
\end{isabelle_cm}
Let us briefly go through the main parts of the formalized proof:
\begin{isabelle_cm}
\isacommand{proof}\ \isacharminus
\end{isabelle_cm}
We set $s, t$ as it is in the algorithm and introduce $I$ the interval for the result:

\begin{isabelle_cm}
\ \ \isacommand{define}\ s\ t\ \isacommand{where}\ s{\isacharunderscore}def\isacharcolon\ s\ =\ {\isasymlceil}8\isacharslash\isasymepsilon{\isacharcircum}2\isasymrceil\ \isacommand{and}\ t{\isacharunderscore}def\isacharcolon\ t\ =\ {\isasymlceil}32\ {\isacharasterisk}\ ln\ {\isacharparenleft}1/\isasymdelta\isacharparenright\isasymlceil\isanewline
\ \ \isacommand{define}\ I\ \isacommand{where}\ I\ =\ \isacharbraceleft\isasymomega\isachardot\ \isacharbar\isasymomega\ \isacharminus\ F2\ xs\isacharbar\ \isasymle\ \isasymepsilon\ \isacharasterisk\ F2\ xs\isacharbraceright\isanewline
\ \ \isacommand{define}\ R\ \isacommand{where}\ R\ h\ \isacharequal\ \isacharparenleft{\isasymSum}j{\isacharless}s\isachardot\ \isacharparenleft{\isasymSum}x\ \isasymleftarrow\ xs\isachardot\ h\ x\isacharparenright\isactrlsup {\isadigit{2}}\isacharparenright{\isacharslash}s
\end{isabelle_cm}

The random variable \isa{R} represents the estimates from \isa{example{\isacharunderscore}5} over the pseudorandom object:
\isa{\isasymH\ \isadigit{2}\ s\ \isacharparenleft\isasymH\ \isadigit{4}\ n\ \isacharparenleft\isasymL\ \isacharbrackleft\isadigit{1}\isacharcomma\isacharminus\isadigit{1}\isacharbrackright\isacharparenright\isacharparenright}.
Based on our previous observation, the random variable \isa{R} will be outside the desired interval with probability at most $\frac{1}{4}$.
\begin{isabelle_cm}
\ \ \isacommand{have}\ measure\ \isacharparenleft\isasymH\ \isadigit{2}\ s\ \isacharparenleft\isasymH\ \isadigit{4}\ n\ \isacharparenleft\isasymL\ \isacharbrackleft\isadigit{1}\isacharcomma\isacharminus\isadigit{1}\isacharbrackright\isacharparenright\isacharparenright\isacharparenright\ {\isacharbraceleft}h\isachardot\ R\ h\ \isasymnotin\ I\isacharbraceright\ \isasymle\ 1{\isacharslash}4
\end{isabelle_cm}
This can be stated as the expectation of an indicator variable:
\begin{isabelle_cm}
\ \ \isacommand{hence}\ \isacharasterisk\isacharcolon{\isasymintegral}h\isachardot\ of{\isacharunderscore}bool{\isacharparenleft}R\ h\ \isasymnotin\ I\isacharparenright\ \isasympartial\isacharparenleft\isasymH\ \isadigit{2}\ s\ \isacharparenleft\isasymH\ \isadigit{4}\ n\ \isacharparenleft\isasymL\ \isacharbrackleft\isadigit{1}\isacharcomma\isacharminus\isadigit{1}\isacharbrackright\isacharparenright\isacharparenright\isacharparenright\ \isasymle\ 1{\isacharslash}4
\end{isabelle_cm}
Let us introduce an abbreviation for the pseudorandom object of \isa{example{\isacharunderscore}7}. It represents sampling using random walks from the pseudorandom object of \isa{example{\isacharunderscore}5}.
\begin{isabelle_cm}
\ \ \isacommand{let}\ {\isacharquery}q\ \isacharequal\ \isasymE\ t\ {\isacharparenleft}1{\isacharslash}8\isacharparenright\ \isacharparenleft\isasymH\ \isadigit{2}\ s\ \isacharparenleft\isasymH\ \isadigit{4}\ n\ \isacharparenleft\isasymL\ \isacharbrackleft\isadigit{1}\isacharcomma\isacharminus\isadigit{1}\isacharbrackright\isacharparenright\isacharparenright\isacharparenright
\end{isabelle_cm}
Now we start an equation chain. First, we can express the left-hand-side as follows: 
\begin{isabelle_cm}
\ \ \isacommand{have}\ {\isacharquery}L\ \isacharequal\ measure\ {\isacharquery}q\ {\isacharbraceleft}h\isachardot\ median\ t\ \isacharparenleft{\isasymlambda}i\isachardot\ R\ {\isacharparenleft}h\ i\isacharparenright\isacharparenright\ \isasymnotin\ I\isacharbraceright 
\end{isabelle_cm}
As mentioned earlier, the median can only be outside the interval if at least half of the estimates are outside:
\begin{isabelle_cm}
\ \ \isacommand{also}\ \isacommand{have} \isachardot\isachardot\isachardot\ \isasymle\ measure\ {\isacharquery}q\ {\isacharbraceleft}h\isachardot\ \isacharparenleft{\isasymSum}j\ \isacharless\ t\isachardot\ of{\isacharunderscore}bool{\isacharparenleft}R\ {\isacharparenleft}h\ j\isacharparenright\ \isasymnotin\ I\isacharparenright\isacharparenright{\isacharslash}t\ \isasymge\ 1{\isacharslash}2\isacharbraceright
\end{isabelle_cm}
The following step is critical. Using the Chernoff bound for expander walks, we can conclude that the probability that half of the estimates will be outside the interval is less than $\exp(-2 l (1/8)^2)$. Here, we used the previous observation \isa{\isacharasterisk}.
\begin{isabelle_cm}
\ \ \isacommand{also}\ \isacommand{have} \isachardot\isachardot\isachardot\ \isasymle\ measure\ {\isacharquery}q\ {\isacharbraceleft}h\isachardot\ \isacharparenleft{\isasymSum}j\ \isacharless\ t\isachardot\ of{\isacharunderscore}bool{\isacharparenleft}R\ {\isacharparenleft}h\ j\isacharparenright\ \isasymnotin\ I\isacharparenright\isacharparenright{\isacharslash}t\ \isacharminus\ 1{\isacharslash}4\ \isasymge\ 1{\isacharslash}8\ \isacharplus\ 1{\isacharslash}8\isacharbraceright\isanewline
\ \ \isacommand{also}\ \isacommand{have} \isachardot\isachardot\isachardot\ \isasymle\ exp\isacharparenleft{\isacharminus}2{\isacharasterisk}t{\isacharasterisk}{\isacharparenleft}1{\isacharslash}8\isacharparenright\isacharcircum2\isacharparenright
\end{isabelle_cm}
The right-hand side is less than $\delta$, which concludes the proof.
\begin{isabelle_cm}
\ \ \isacommand{also}\ \isacommand{have} \isachardot\isachardot\isachardot\ \isasymle\ \isasymdelta\isanewline
\ \ \isacommand{finally}\ \isacommand{show}\ {\isacharquery}thesis\ \isacommand{by}\ simp\isanewline
\isacommand{qed}
\end{isabelle_cm}
Note that the resulting algorithm has a space complexity of $\bigo(\ln n + \varepsilon^{-2} \ln(\delta^{-1}) \ln m)$%
 for $n, m, \delta^{-1}, \varepsilon^{-1} \rightarrow \infty$.

In 1999, Alon \etal~\cite[\S 2.2]{alon1999} presented a solution for the estimation of the second frequency moment using $\bigo(\varepsilon^{-2} \ln(\delta^{-1}) (\ln n + \ln m))$.
That solution was relying only on a single hash-family.
This section demonstrates an asymptotically more space-efficient algorithm, and its construction was possible simply by using additional pseudorandom objects.
An optimal solution, which also relies on $k$-independent hash families, for the above problem has been found by Kane \etal~\cite{kane2010_2}, but only for a constant failure probability $\delta$.
Their solution uses $\bigo(\ln \ln n + \varepsilon^{-2} \ln m)$ bits, and even works for any (possibly fractional) frequency moment $F_k$ with $0 < k \leq 2$.
It is possible to adapt their algorithm using the previously mentioned median technique to achieve $\bigo(\ln(\delta^{-1}) \ln \ln n + \ln(\delta^{-1}) \varepsilon^{-2} \ln m)$ for non-constant $\delta$.
As far as I know, an optimal solution with respect to all parameters $\varepsilon, \delta, n, m$ is not known. 
Note that, it depends on the magnitude of $n, \delta$, whether the algorithm we developed in this section is preferable to the solution by Kane \etal

The above was a minimal example to demonstrate how the library works and how combining pseudorandom objects can be used to improve a randomized algorithm.
In 2023, I had presented the formalization of a space-optimal algorithm for the distinct elements problem~\cite{karayel2023, Distributed_Distinct_Elements-AFP, karayel2023arxiv}.
Since we are familiar with the notation, it makes sense to replicate the pseudorandom object I used there:

\begin{isabelle_cm}
  \isasymE\ m\ \isasymalpha\ \isacharparenleft\isasymE\ l\ \isasymLambda\ \isacharparenleft \isasymH\ \isadigit{2}\ n \isacharparenleft\isasymG\ n{\isacharunderscore}exp\isacharparenright\ \isasymtimes\isactrlsub{P}\ \isasymH\ \isadigit{2}\ n\ \isacharparenleft\isasymN\ {\isacharparenleft}C\isactrlsub{7}{\isacharasterisk}b\isactrlsup{2}\isacharparenright\isacharparenright\ \isasymtimes\isactrlsub{P}\ \isasymH\ k\ {\isacharparenleft}C\isactrlsub{7}{\isacharasterisk}b\isactrlsup{2}\isacharparenright\ \isacharparenleft\isasymN\ b\isacharparenright\isacharparenright\isacharparenright
\end{isabelle_cm}

A detailed explanation of all the parameters can be found in the source material~\cite{karayel2023arxiv}. However, this highlights that, depending on the algorithm, very different
constructions may be required and points out the importance of the flexibility of the library. 
The above uses the pseudorandom primitive \isa{\isasymG\ n}, which was not mentioned before. This primitive has a geometric distribution%
, \ie, the probability of the value $0$ is $\frac{1}{2}$, the probability of the value $1$ is $\frac{1}{4}$ \textit{etc}.

\section{Conclusion\label{sec:concl}}
This work presents pseudorandom objects as versatile components in randomized algorithms that can improve the asymptotic space and randomness complexity of randomized algorithms.
We discussed the mathematical foundations and the formal verification of two such objects in depth: expander graphs and $k$-independent hash families. 

Novel in this work is the viewpoint that they can be represented as a composable library that allows the construction of application-specific pseudorandom objects from primitive ones.
The notation introduced by this work allows the concise presentation of these algorithms informally and formally in Isabelle/HOL.
The high-level theorems provided by the library allow small and easy proofs (even if complex pseudorandom objects are needed).

Similarly, the formalization of a theory of efficient non-prime finite fields, or expander graphs has---as far as I know---not been done before.

The formalization of the space-optimal distinct elements problem~\cite{karayel2023,karayel2023arxiv} highlights how this framework helps tackle advanced and complex randomized algorithms (that required many decades of research to develop.)

While this work has been done in Isabelle/HOL~\cite{nipkow2002}, the approach is transferrable to other theorem provers/checkers/proof assistants such as Agda, Rocq, Idris, Lean, etc. However, depending on the proof assistant, this may require the development of additional libraries for algebra, analysis, and probability theory.

There is still much work that could be done in this field. In the context of expander graphs, there are still important constructions that could be formalized.
Another task is formalizing Chernoff bounds for random walks in expander graphs for non-discrete functions~\cite{gillman1998,lezaud1998,rao2017}.
Beyond that, further pseudorandom objects%
, such as randomness extractors and list-decodable codes, are open formalization tasks.

\section*{Acknowledgements}
I would like to thank Tobias Nipkow, Yong Kiam Tan and the anonymous reviewers for feedback and many insightful suggestions and remarks on earlier versions of this work.

\printbibliography

\appendix
\newpage
\section{AFP entries\label{apx:form}}
The following list summarizes the AFP entries that contain the results and definitions for pseudorandom objects, with a total of 16365 LoC.
\begin{description}[style=nextline,leftmargin=8pt]
\item[Finite Fields~\cite{Finite_Fields-AFP}]
Formalization of finite fields and algorithms for efficient computations in and constructions of them. (7173 LoC)
\item[Interpolation Polynomials (in HOL-Algebra)~\cite{Interpolation_Polynomials_HOL_Algebra-AFP}]
Existence and count of interpolation polynomials for a given set of points and degree bound. This is used in the construction of hash families. (712 LoC)
\item[Universal Hash Families~\cite{Universal_Hash_Families-AFP}]
Formalization and construction of $k$-independent hash families as pseudrandom objects. Section~\ref{sec:hash} summarizes this AFP entry. (1435 LoC)
\item[Expander Graphs~\cite{Expander_Graphs-AFP}]
Formalization and construction of (strongly explicit) expander graphs. Section~\ref{sec:expander} summarizes this AFP entry. (7054 LoC)
\end{description}
The following list summarizes some AFP entries relevant for the verification of randomized algorithms:
\begin{description}[style=nextline,leftmargin=8pt]
\item[Median Method~\cite{Median_Method-AFP}]
Concept of the median (and order statistics in general) and their properties as a random variable. It verifies that order statistics are measurable.
In particular, it contains the results about the median, we used in Section~\ref{sec:combining_pros}. (484 LoC)
\item[Concentration Inequalities~\cite{Concentration_Inequalities-AFP}]
\emph{Joint work with Yong Kiam Tan.} Collection of advanced concentration inequalities, such as the Efron--Stein or Paley--Zygmund inequality. Includes also Bienaym\'{e}'s identity, which we used in Section~\ref{sec:combining_pros}. (2019 LoC)
\end{description}
The following entries are verifications of randomized algorithms that rely on pseudorandom objects:
\begin{description}[style=nextline,leftmargin=8pt]
\item[Distributed Distinct Elements~\cite{Distributed_Distinct_Elements-AFP}]
Verification of a space-optimal cardinality estimation algorithm~\cite{karayel2023, karayel2023arxiv}. (4981 LoC)
\item[Formalization of Randomized Approximation Alg. for Frequency Moments~\cite{Frequency_Moments-AFP}]
Verification of several approximation algorithms for frequency moments on streams~\cite{karayel2022}. (3791 LoC)
\item[Approximate Model Counting~\cite{Approximate_Model_Counting-AFP}]
\emph{By Yong Kiam Tan and Jiong Yang.} Verification of a randomized approximation algorithm for the count of satisfying assignments to a boolean formula~\cite{yongkiam2024}.
It relies on the construction of a $3$-universal XOR-hash-family. (See also \cite[RandomXORHashFamily]{Approximate_Model_Counting-AFP}.)
 (5890 LoC)
\end{description}

\end{document}